\documentclass{siamart171218}
\usepackage[margin=0.5in]{geometry}

\usepackage{amsmath}
\usepackage{amsfonts}
\usepackage{graphicx}
\usepackage{caption}
\usepackage{subcaption}
\usepackage{color}
\usepackage{esint}

\usepackage[T1]{fontenc}

\begin{document}

\title{The Stability and Dynamics of Localized Spot Patterns for a Bulk-Membrane Coupled Brusselator Model}

\author{Daniel Gomez\thanks{Dept. of Mathematics, UBC, Vancouver,
		Canada (\email{dagubc@math.ubc.ca})}} \baselineskip=16pt

\baselineskip=16pt

\maketitle

\begin{abstract}
We consider a bulk-membrane-coupled partial differential equation in which a single diffusion equation posed within the unit ball is coupled to a two-component reaction diffusion equation posed on the bounding unit sphere through a linear Robin boundary condition. Specifically, within the bulk we consider a process of linear diffusion with point-source generation for a bulk-bound activator. On the bounding surface we consider the classical two-component Brusselator model where the feed term is replaced by the restriction of the bulk-bound activator to the membrane. By considering the singularly perturbed limit of a small diffusivity ratio between the membrane-bound activator and inhibitor species, we use formal asymptotic expansions to construct strongly localized quasi-equilibrium spot solutions and study their linear stability. Our analysis reveals that bulk-membrane-coupling can restrict the existence of localized spot solutions through a recirculation mechanism. In addition we derive stability thresholds that illustrate the effect of coupling on both competition and splitting instabilities. Finally, we use higher-order matched asymptotic expansions to derive a system of differential algebraic equations that describe the slow motion of spots. The potential for new coupling induced dynamical behaviour is illustrated by considering examples of one-, two-, and three-spot solutions. 
\end{abstract}

\section{Introduction}\label{sec:introduction}

A central problem in the study of early developmental biology is to
both determine the mechanisms driving structural changes and to then
describe the patterned structures that emerge. In one proposed
mechanism a collection of chemicals collectively known as
\textit{morphogens} diffuse and react with each other leading to a
concentration distribution, known as a \textit{prepattern}, that
serves as a template for later structural changes. Although
experimental evidence of morphogens remains absent these models have
been successful in qualitatively generating patterns readily found in
biological systems. These models are mathematically described by
systems of reaction diffusion (RD) equations to be solved for the
morphogen concentrations. The first steps forward in this theory can
be traced back to the pioneering work of Alan M. Turing
\cite{turing_1952} in which he demonstrated that under certain
conditions on the species' diffusivities, spatially homogeneous
solutions to a two-species RD system can bifurcate to spatially
heterogeneous solutions. This idea has since spurred an
immense body of literature analysing these \textit{Turing
  instabilities} for RD systems having different prescribed
kinetics (see for example the review article by Maini et.\@ al.\@ \cite{maini2012} as well as the textbook by Murray \cite{murray2003}).

One shortcoming of the morphogen prepattern theory is that the
criteria for Turing instabilities to be triggered may require
unrealistically large differences between the chemical species'
diffusivities. Recently, a growing body of literature has avoided this
limitation by proposing models that couple RD systems posed within a
cell's bulk (or cytosol) to RD systems posed on the cell membrane. In
this class of \textit{bulk-membrane coupled reaction diffusion
  systems} a biologically motivated assumption is that the membrane
diffusivities are typically much smaller than their cytosol
counterparts \cite{levine_2005}. Using a combination of linear
stability analysis and numerical experimentation several studies have
determined that by introducing bulk-membrane coupling it is possible
to trigger Turing instabilities within the bulk or membrane in
parameter regimes where the isolated uncoupled systems would not
exhibit such behaviour
\cite{madzvamuse_2015,madzvamuse_2016,ratz_2012,ratz_2013,ratz_2014}. In
addition, bulk-membrane coupled models have also been used to
successfully describe a possible cell-polarizing mechanism in which
rather than a Turing instability, the driving mechanism is a
competition between mass conservation and an autocatalytic reaction of
a single chemical species \cite{diegmiller_2018,cusseddu2019}.

\begin{figure}[t!]
\centering
\includegraphics[scale=0.75]{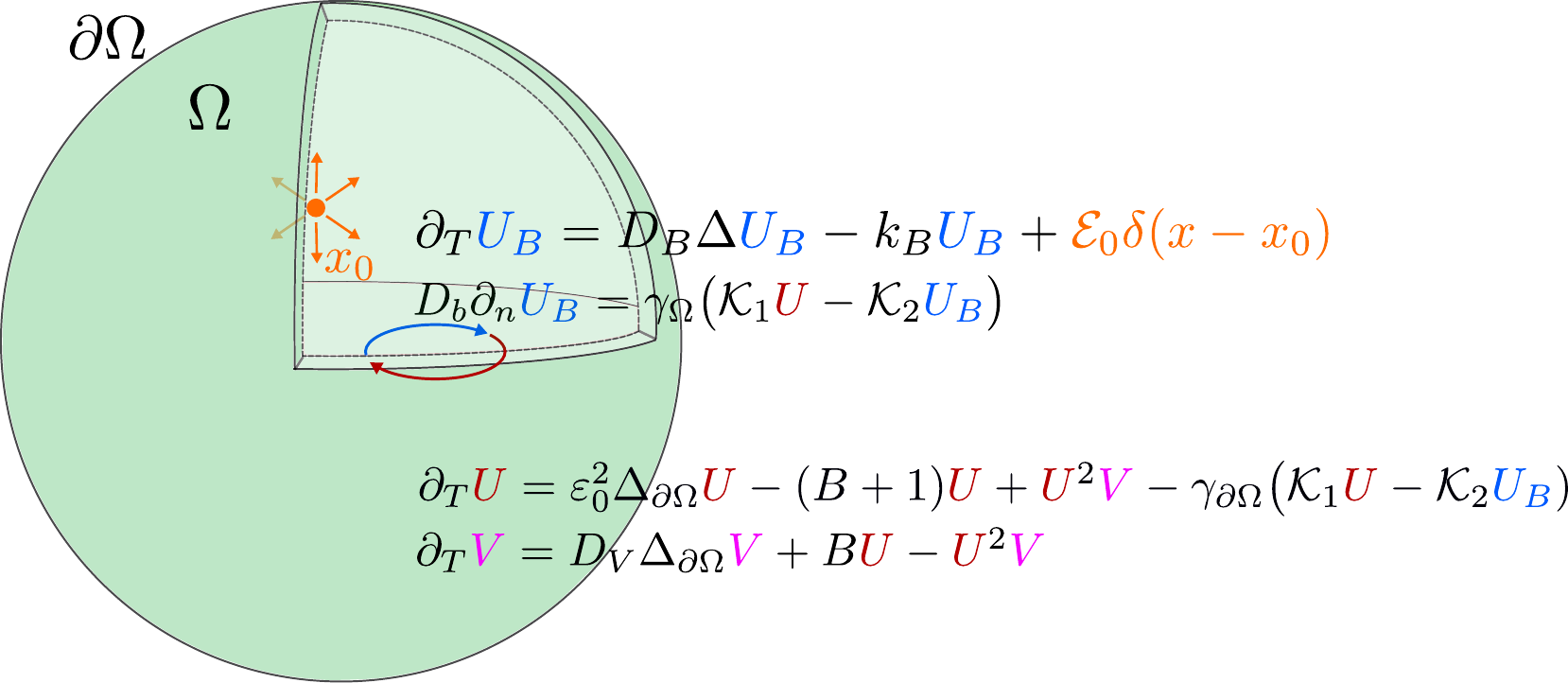}
\caption{Schematic plot illustrating the geometry of the bulk-membrane coupled model being considered.}\label{fig:schematic}
\end{figure}

In this paper we consider a bulk-membrane coupling extension to the
previously studied problem of a singularly perturbed Brusselator RD
system posed on the unit sphere. Using techniques from singular
perturbation theory the authors in \cite{rozada2014} asymptotically
constructed a quasi-equilibrium solution consisting of $N$ spots,
corresponding to regions where the activator is strongly localized,
arranged on the unit sphere. Additionally, their analysis revealed
that these $N$-spot patterns are susceptible to instabilities in
$O(1)$ time that lead to spots splitting and replicating, or competing
and annihilating each other. This work was then extended by Trinh and
Ward \cite{trinh2016} to account for the long-time behaviour
of $N$-spot patterns which they determined is governed by a
system of differential algebraic equations (DAE) in the spot locations. Our primary
goal will therefore be to analyse the effect that bulk-membrane
coupling has on spot splitting and competition instabilities, as well as
on the slow spot dynamics of $N$-spot patterns for the Brusselator
model. Specifically, we consider a reaction-diffusion system with
Brusselator kinetics posed on the unit sphere, coupled to a bulk
linear diffusion process for the activator within the unit ball. With
$\Omega$ being the unit ball in $\mathbb{R}^3$ we consider the
reaction diffusion system
\begin{subequations}\label{eq:pde-unscaled}
\begin{align}
& \partial_T U = \varepsilon_0^2\Delta_{\partial\Omega} U - (B+1) U + U^2 V - \gamma_{\partial\Omega}\bigl(\mathcal{K}_1 U - \mathcal{K}_2 U_B), &\text{in }\partial\Omega, \label{eq:pde-unscaled-U} \\
& \partial_TV = D_V\Delta_{\partial\Omega} V + B U - U^2 V, &\text{in }\partial\Omega, \label{eq:pde-unscaled-V}
\end{align}
\end{subequations}
for the membrane-bound activator and inhibitor concentrations $U(x,T)$
and $V(x,T)$ respectively, coupled to a diffusion equation within the
bulk
\begin{subequations}
\begin{align}\label{eq:pde-unscaled-UB}
& \partial_TU_B = D_B\Delta U_B - k_B U_B + \mathcal{E}_0 \delta(x-x_0), & \text{in }\Omega,\\
& D_b\partial_n U_B = \gamma_{\Omega}\bigl( \mathcal{K}_1 U - \mathcal{K}_2 U_B\bigr), & \text{on }\partial\Omega,
\end{align}
\end{subequations}
for the bulk-bound activator $U_b(x,T)$. A schematic representation is shown in Figure \ref{fig:schematic}. We remark that recent studies considering only the membrane-bound Brusselator model have been used to model conifer morphogenesis \cite{holloway2018,charette2018}. By introducing bulk-membrane coupling our model gives a clear origin to the feed term typically found in the membrane-bound activator equation for uncoupled membrane-bound Brusselator models. Specifically, the bulk-bound equation \eqref{eq:pde-unscaled-UB} describes a site of ongoing activator generation of strength $\mathcal{E}_0$ concentrated at a point $x_0$ within the bulk. The bulk-bound bound activator generated in this way then diffuses and attaches to the membrane where it provides the necessary feed term required to sustain the formation of patterns.

In Appendix \ref{app:scaling} we introduce an appropriate scaling so
that \eqref{eq:pde-unscaled} exhibits localized spot patterns. The
non-dimensionalized problem is then given by a system of RD equations posed
on the membrane
\begin{subequations}\label{eq:pde-all}
\begin{align}
u_t & = \varepsilon^2 \Delta_{\partial\Omega} u - u + f u^2 v + \varepsilon^2 K_2 w|_{\partial\Omega}, & \text{in}\quad \partial\Omega, \label{eq:pde-u}\\ 
\tau_v v_t & = D_v\Delta_{\partial\Omega} v + \varepsilon^{-2}(u - u^2 v), & \text{in}\quad \partial\Omega, \label{eq:pde-v}
\end{align}
which are coupled to a single diffusion equation with feed term defined inside the bulk
\begin{equation}
\tau_w w_t = D_w \Delta w - w  + E_0\delta(x-x_0),\quad\text{in}\quad\Omega,\qquad D_w \partial_n  w + K_2 w = \varepsilon^{-2}K_1 u,\quad\text{on}\quad\partial\Omega. \label{eq:pde-w}
\end{equation}
\end{subequations}
We remark that the scaling in Appendix \ref{app:scaling} naturally leads to the parameter constraints
$$
0<f<1,\quad 0<\varepsilon\ll 1, \quad 0\leq K_1 < 1,\quad 0\leq K_2 < \infty.
$$
In \eqref{eq:pde-v} and \eqref{eq:pde-u}, the
Laplace-Beltrami operator $\Delta_{\partial\Omega}$ in spherical
coordinates $(\theta,\varphi)$ has the form
\begin{equation}\label{eq:laplace-beltrami}
\Delta_{\partial\Omega} = \frac{1}{\sin\theta}\partial_\theta\sin\theta\partial_\theta +  \frac{1}{\sin^2\theta}\partial_\varphi^2.
\end{equation}

The introduction of bulk-membrane coupling to the classical
Brusselator model has two novel features which are best illustrated by
using the linearity of \eqref{eq:pde-w} to write
$$
w(x,t) = U(x,t) + E(x),
$$
where $U(x,t)$ satisfies
$$
\tau_w U_t = D_w \Delta U - U,\quad\text{in}\quad\Omega,\qquad D_w\partial_n U + K_2 U = \varepsilon^{-2}K_1 u,\quad\text{on}\quad\partial\Omega,
$$
while $E(x)$ satisfies the time-independent problem
$$
D_w \Delta E - E = - E_0\delta(x-x_0),\quad\text{in}\quad\Omega,\qquad D_w\partial_n E + K_2 E = 0,\quad\text{on}\quad\partial\Omega.
$$
Therefore $E(x)$ describes the concentration of a diffusing point
source located within the bulk, while $U(x,t)$ describes the diffusion
of $u$ into the bulk through a Langmuir-type boundary condition. The
term $E(x)$ within the membrane equation \eqref{eq:pde-u} serves as a
substitute for the typical source term needed to sustain patterns in
the Brusselator model. This source term will be spatially-homogeneous
(resp. heterogeneous) if $\eta_0=0$ (resp. $0<\eta_0<1$) and the
effects of heterogeneous sources have previously been studied for
two-dimensional domains \cite{tzou2018}. Thus any results linked to
$E(x)$ are not the product of bulk-membrane coupling but
rather of heterogeneity. In contrast, the effect of $U(x,t)$ is a
direct reflection of the bulk-membrane coupling. Indeed, our analysis
reveals that $U(x,t)$ can be interpreted as a recirculation of the
membrane-bound activator through the bulk, having direct consequences
on both the existence of localized spot solutions as well on their
stability. Our analysis of the slow ODE dynamics further reveals that
recirculation may lead to novel asymmetric spots.

This paper is organized as follows. In Section
\ref{sec:quasi-equilibrium} we use the method of matched asymptotic
expansions to construct quasi-equilibrium $N$-spot configurations that
are stationary on an $O(1)$ time-scale. Our analysis reveals that
coupling plays a key role in the existence of such spots given our
scaling regime. In Section \ref{sec:stability} we consider the linear
stability of the $N$-spot quasi-equilibrium configurations on an $O(1)$
time-scale. Our analysis focuses on spitting and competition
instabilities. The linearized system is known to exhibit
asymptotically small eigenvalues which correspond to drift
instabilities on a long, $O(\varepsilon^{-2})$, time scale and we
address this in Section \ref{sec:dynamics} where we derive the
relevant differential algebraic system describing slow spot motion. We
then consider examples of one-, two-, and three-spot configurations in
Section \ref{sec:examples}. For two-spot configuration we carry out a
detailed application of the theory developed in Sections
\ref{sec:stability} and \ref{sec:dynamics} to determine regions where
two-spot configurations are stable on both $O(1)$ and
$O(\varepsilon^{-2})$ time scales. Finally, in Section
\ref{sec:discussion} we summarize our results and point to several
directions for future research.

\section{Asymptotic Construction of Quasi-Equilibria}\label{sec:quasi-equilibrium}

In this section we use the method of matched asymptotic expansions to construct quasi-equilibrium solutions to \eqref{eq:pde-all} consisting of $N$ localized \textit{spots} arranged on the membrane at
\begin{equation}\label{eq:x_i}
x_i = (\sin\theta_i\cos\varphi_i, \sin\theta_i\sin\varphi_i,\cos\theta_i)^T,\qquad i=1,...,N,
\end{equation}
with the separation constraints
$$
|x_i - x_j| = O(1)\qquad \forall\quad i\neq j,\qquad 1 - \eta_0 = O(1).
$$
The method of matched asymptotic expansions has been successfully
employed for a wide variety of singularly perturbed problems, with
those most pertinent to us being the stability analysis for the
(uncoupled) Brusselator on the sphere in \cite{rozada2014} and the
derivation of the ODE system describing slow-spot dynamics in
\cite{trinh2016}. A key step in the proceeding analysis is the
introduction of local coordinates near the $i^\text{th}$ spot, for
which both \cite{rozada2014} and \cite{trinh2016} used
$$
y_1 := \sin\theta_i\frac{\varphi - \varphi_i}{\varepsilon},\qquad y_2 := \frac{\theta - \theta_i}{\varepsilon}.
$$
With this choice the Laplace-Beltrami operator $\Delta_{\partial\Omega}$ becomes
$$
\varepsilon^2 \Delta_{\partial\Omega} = \Delta_y + \varepsilon\cot\theta_i(\partial_{y_2} - 2y_2\partial_{y_1}^2) + O(\varepsilon^2),\qquad \Delta_y := \partial_{y_1}^2 + \partial_{y_2}^2.
$$
To construct quasi-equilibria and study their $O(1)$ stability only
the leading order term is needed. However, to derive an ODE system
governing the spots' long time dynamics we must also use the
$O(\varepsilon)$ correction. As highlighted in \cite{trinh2016} this
leads to a sub-problem that is explicitly solvable regardless of the
model being used and in this sense can be interpreted as an artifact
of the choice of local coordinates. Indeed we see that the
$\cot\theta_i$ term appearing in the $O(\varepsilon)$ correction
introduces a $\theta_i$ dependence to the local problem near the
$i^{\text{th}}$ spot, in conflict with the sphere's symmetry. In
order to bypass these artificial effects we use choose $(y_1,y_2)$ to
be stretched \textit{geodesic normal coordinates} for which (see
Appendix A of \cite{tse2010}) the Laplace-Beltrami operator becomes
\begin{equation}\label{eq:local-laplace-beltrami}
\varepsilon^2 \Delta_{\partial\Omega} = \Delta_y + O(\varepsilon^2),\qquad\qquad \Delta_y := \partial_{y_1}^2 + \partial_{y_2}^2.
\end{equation}

To explicitly construct the stretched normal coordinates $(y_1,y_2)$ on the sphere at $x_i$ we first remark that spherical coordinates are already normal coordinates along the equator $\theta = \pi/2$. Next we let $R_i$ be the rotation taking $x_i$ to $(1,0,0)^T$ given by
\begin{equation*}
R_i := \begin{pmatrix} \sin\theta_i & 0 & \cos\theta_i \\ 0 & 1 & 0 \\ -\cos\theta_i & 0 & \sin \theta_i\end{pmatrix}\begin{pmatrix} \cos\varphi_i &  \sin\varphi_i & 0 \\ -\sin\varphi_i & \cos\varphi_i & 0 \\ 0 & 0 & 1\end{pmatrix}.
\end{equation*}
Introducing spherical coordinates $(\tilde{\theta},\tilde{\varphi})$
in the rotated frame $\tilde{x} = R_i x$ we define the stretched local
coordinates at $x_i$ given by
\begin{equation}\label{eq:local-coordinates}
y_1 = \varepsilon^{-1}\tilde{\varphi},\qquad y_2 = \varepsilon^{-1}\biggl(\tilde{\theta} - \frac{\pi}{2}\biggr),
\end{equation}
in terms of which we have the local expansion
\eqref{eq:local-laplace-beltrami} for the Laplace-Beltrami
operator. To perform the method of matched asymptotics we develop the
following formulas relating the inner variables $y$ near each spot,
with the outer variables $x\in\Omega\cup\partial\Omega$. We have for
$\varepsilon\ll 1$ and $x$ near $x_i$, the expansion
\begin{equation}\label{eq:x-y-relationship-1}
x - x_i = R_i^{T}(\tilde{x} - \tilde{x}_i) = \varepsilon R_i^T \begin{pmatrix} 0 \\ y_1 \\ -y_2 \end{pmatrix} -\frac{1}{2}\varepsilon^2\rho^2 R_i^ T\begin{pmatrix} 1 \\ 0 \\ 0 \end{pmatrix} +   O(\varepsilon^3) = \varepsilon \mathcal{J}_i y - \frac{1}{2}\varepsilon^2 \rho^2 x_i + O(\varepsilon^3),
\end{equation}
where
\begin{equation}\label{eq:J_i-definitionJ}
\mathcal{J}_i = \begin{pmatrix} -\sin\varphi_i & \cos\theta_i\cos\varphi_i \\ \cos\varphi_i & \cos\theta_i\sin\varphi_i \\ 0 & -\sin\theta_i \end{pmatrix},
\end{equation}
and we remark that
\begin{equation}\label{eq:J_i_identities}
\mathcal{J}_i^T\mathcal{J}_i = \mathbb{I}_2, \qquad\text{and}\quad \mathcal{J}_i \mathcal{J}_i^T = \mathbb{I}_3 - x_i x_i^T,
\end{equation}
where $\mathbb{I}_d$ denotes the identity matrix in
$d$-dimensions. Since $\mathcal{J}_i^T x_i = 0$ we calculate for $|x-x_i|=O(\varepsilon)$
$$
|x-x_i|^2 = \varepsilon^2\bigl( y^T \mathcal{J}_i^T\mathcal{J}_i y - \frac{1}{2}\varepsilon\rho^2 y^T \mathcal{J}_i^T x_i - \frac{1}{2}\varepsilon\rho^2 x_i^T \mathcal{J}_i y  + O(\varepsilon^2)\bigr) = \varepsilon^2 \rho^2 + O(\varepsilon^4),
$$
and therefore
\begin{equation}\label{eq:local-global-norm-x-x_i}
|x-x_i| = \varepsilon\rho + O(\varepsilon^3).
\end{equation}
Next, for $|\xi - x_i| = O(1)$ but $|x-x_i|=O(\varepsilon)$ we calculate
\begin{equation}\label{eq:local-global-norm-x-x_i-xi}
|x-\xi|^2 = |x_i-\xi|^2\biggl( 1 - \frac{2 \varepsilon y^T \mathcal{J}_i^T(\xi - x_i)}{|\xi-x_i|^2} + O(\varepsilon^2)   \biggr)
\end{equation}

Using the local normal coordinates introduced above we now proceed
with the asymptotic matching. We begin by introducing the
following asymptotic expansions for $|x-x_i| = O(\varepsilon)$
\begin{equation*}
u \sim D_v^{1/2}u_{i0}(y) + o(1),\qquad v\sim D_v^{-1/2}v_{i0}(y) + o (1),\qquad w \sim \frac{K_1 D_v^{1/2}}{\varepsilon D_w} w_{i0}(y,y_3) + o(\varepsilon^{-1}),
\end{equation*}
where $y_3 = \varepsilon^{-1}(1-r)$. It follows that the membrane
bound species are given by radially symmetric solutions to the
familiar core-problem
\begin{subequations}\label{eq:core-problem-membrane}
\begin{align}
&\Delta_\rho u_{i0} - u_{i0} + f u_{i0}^2 v_{i0} = 0,\qquad \Delta_\rho v_{i0} + u_{i0} - u_{i0}^2 v_{i0} = 0,\qquad \text{in}\quad\rho >0, \label{eq:core-problem-pde}\\
&u_{i0}'(0) = v_{i0}'(0) = 0,\quad \text{and}\quad u_{i0}\rightarrow 0,\quad v_{i0} \sim S_i \log \rho + \chi(S_i,f)\quad\text{as }\rho\rightarrow\infty,
\end{align}
\end{subequations}
where
\begin{equation*}
\rho := \sqrt{y_1^2 + y_2^2}\quad\text{and}\quad \Delta_\rho := \frac{\partial^2}{\partial\rho^2} + \frac{1}{\rho}\frac{\partial}{\partial\rho},
\end{equation*}
while the bulk-bound activator satisfies
\begin{equation*}
(\Delta_y + \partial_{y_3}^2) w_{i0} = 0,\quad\text{in }y\in\mathbb{R}^2, y_3 > 0,\qquad -\partial_{y_3}w_{i0} = u_{i0}(\rho),\qquad\text{on }y\in \mathbb{R}^2, y_3=0.
\end{equation*}

\begin{figure}[t!]
	\centering
	\begin{subfigure}{0.5\textwidth}
		\centering
		\includegraphics[scale=0.75]{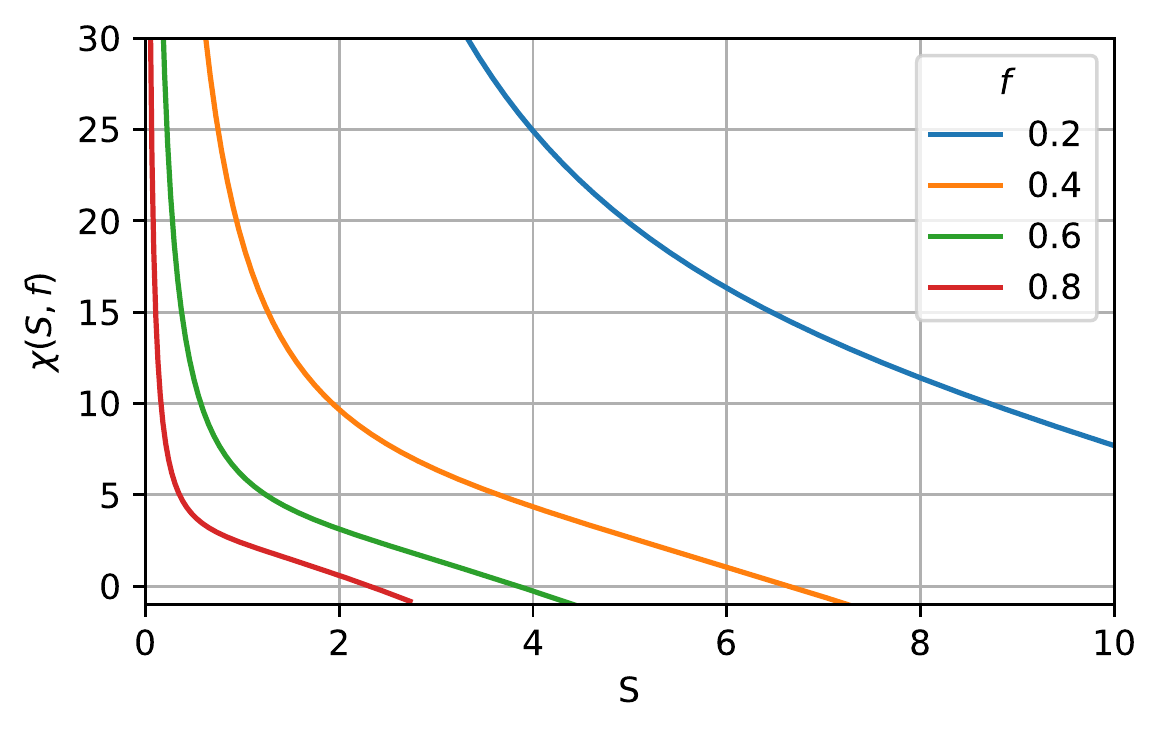}
		\caption{}\label{fig:chi_plots}
	\end{subfigure}%
	\begin{subfigure}{0.5\textwidth}
		\centering
		\includegraphics[scale=0.75]{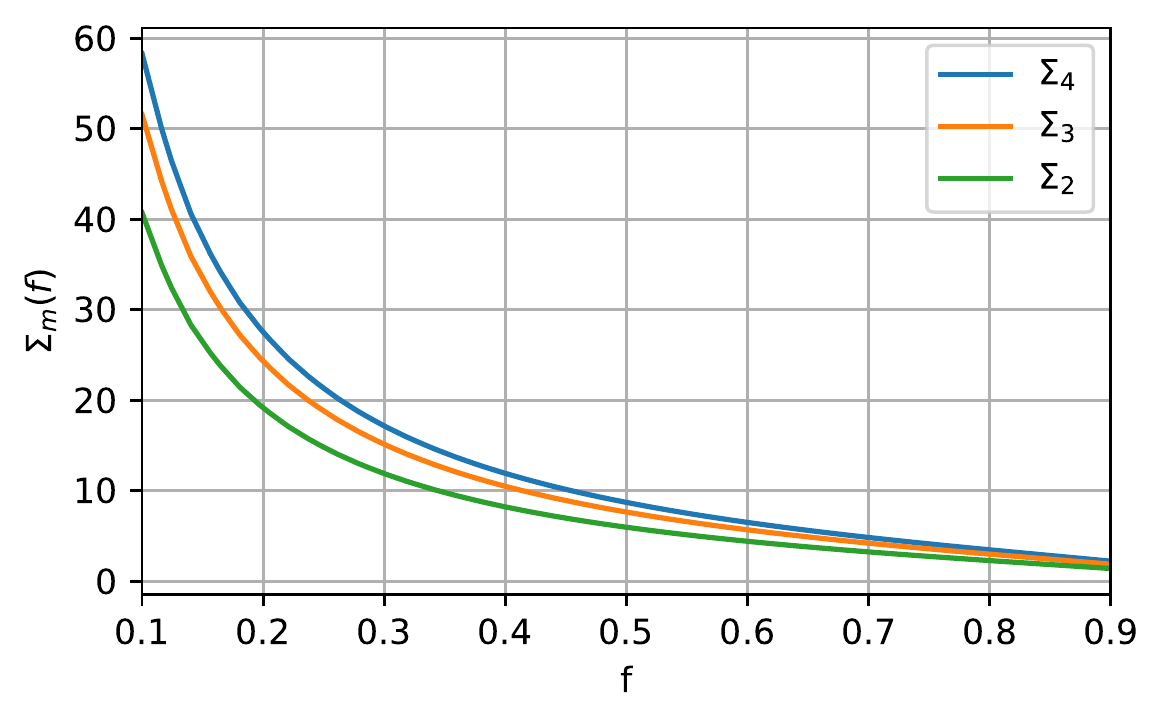}
		\caption{}\label{fig:splitting_instability}
	\end{subfigure}%
	\caption{Plots of (a) the constant in the far-field  behaviour $v_{0}\sim S\log\rho + \chi(S;f)$ for $\rho\rightarrow\infty$ as $S$ is varied for fixed values of $f$, and (b) the $m\geq 2$ instability thresholds $S = \Sigma_m(f)$ versus $f$ for fixed values of the mode $m$.}\label{fig:chi_and_sigma_plots}
\end{figure}

The membrane-bound core-problem \eqref{eq:core-problem-membrane} is
identical to that encountered in previous studies on the sphere
\cite{rozada2014,trinh2016}. We
can numerically solve for $u_{i0}$ and $v_{i0}$ after specifying $f$
and $S_i$. The values of $S_i$ are currently unknown, and will be
found as solutions to a nonlinear system of equations obtained by
matching to the outer solution. It is crucial therefore to solve for
$\chi(S,f)$ and this is done numerically, with sample curves for fixed
values of the parameter $f$ being shown in Figure
\ref{fig:chi_plots}. Applying the divergence theorem to
$u_{i0} + f v_{i0}$ we also obtain the useful relationship
\begin{equation}\label{eq:S_i-integral}
S_i = \frac{1-f}{f}\int_0^\infty u_{i0}(\rho)\rho d\rho > 0.
\end{equation}
Once $u_{i0}$ and $v_{i0}$ have been determined we then easily calculate
\begin{equation*}
w_{i0}(y_1,y_2,y_3) = \frac{1}{2\pi}\int_{-\infty}^{\infty}\int_{-\infty}^{\infty} \frac{u_{i0}(\sqrt{\xi_1^2+\xi_2^2})}{\sqrt{(\xi_1-y_1)^2 + (\xi_2-y_2)^2 + y_3^2}}d\xi_1 d\xi_2 + C,
\end{equation*}
where $C$ is an undetermined constant. The two key properties of
$w_{i0}$ are that it is bounded and that it is radially symmetric on
the plane $y_3=0$.

\begin{figure}[t!]
\centering
\includegraphics[scale=0.8]{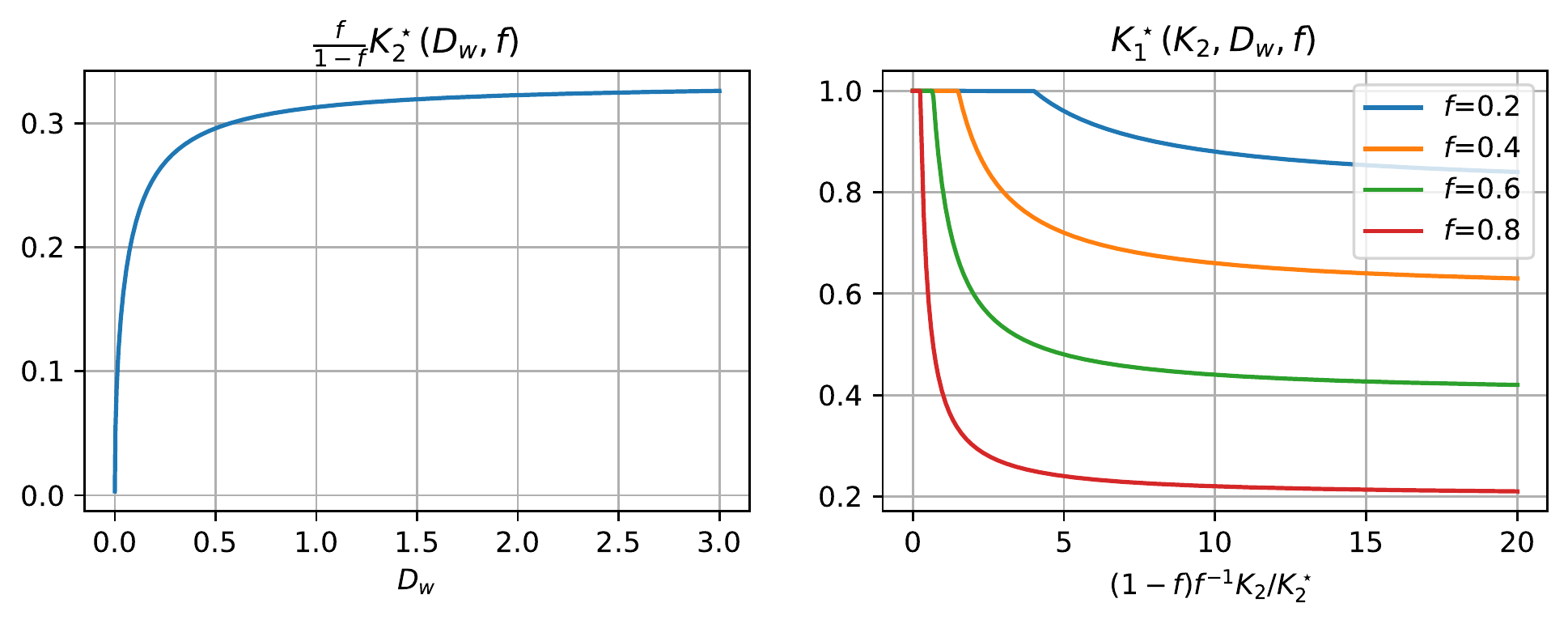}
\caption{Parameter dependence of the existence threshold. Localized
  spot patterns are predicted to exist only when $K_1$ lies beneath
  the curves in the right
  figure.}\label{fig:existence-threshold-illustration}
\end{figure}

Using \eqref{eq:local-global-norm-x-x_i} and the exponential decay of each $u_{i0}(\rho)$ we obtain the leading order approximation
$$
u(x) \sim \varepsilon^2 K_2 w\bigr|_{\partial\Omega} + D_v^{1/2}\sum_{i=1}^N u_{i0}\biggl( \frac{|x-x_i|}{\varepsilon} \biggr),
$$
for the membrane-bound activator. Taking $\varepsilon\rightarrow 0^+$ and using \eqref{eq:S_i-integral} we obtain, in the sense of distributions, the limits
$$
\frac{u-u^2v}{\varepsilon} \sim K_2 w(x) - 2\pi D_v^{1/2}\sum_{i=1}^N S_i \delta_{\partial\Omega}(x-x_i),\qquad \frac{u}{\varepsilon^2} \sim K_2 w(x) + \frac{2\pi f D_v^{1/2}}{1-f}\sum_{i=1}^N S_i\delta_{\partial\Omega}(x-x_i).
$$
The outer solution, valid for $|x-x_i|=O(1)$ ($i=1,...,N$), is thus found by solving
\begin{equation}\label{eq:pde-outer-v}
D_v \Delta_{\partial\Omega} v = -K_2 w + 2\pi D_v^{1/2}\sum_{i=1}^N S_i\delta_{\partial\Omega}(x-x_i),\qquad\text{in }\partial\Omega,
\end{equation}
and
\begin{subequations}\label{eq:pde-outer-w}
\begin{align}
&D_w\Delta w - w = -E_0\delta(x-x_0), & \text{in }\Omega, \\
&D_w\partial_nw + K_2(1-K_1) w = \frac{2\pi f K_1 D_v^{1/2}}{1-f}\sum_{i=1}^N S_i \delta_{\partial\Omega}(x-x_i), & \text{on }\partial\Omega.
\end{align}
\end{subequations}
We introduce the membrane Green's function $G_m(x,\xi)$ with
$x,\xi\in\partial\Omega$ that satisfies
\begin{equation*}
\Delta_{\partial\Omega} G_m = \frac{1}{|\partial\Omega|} -\delta_{\partial\Omega}(x-\xi),\quad\text{in }\partial\Omega,\qquad \int_{\partial\Omega} G_m dA = 0.
\end{equation*}
We also introduce two Robin Green's functions, $G_{rb}(x,\xi)$ and
$G_{rm}(x,\xi)$, where the first has a bulk-bound source term
$\xi\in\Omega$ and satisfies
\begin{equation*}
\Delta G_{rb} - \mu^2 G_{rb} = -\delta(x-\xi),\quad\text{in }\Omega,\qquad \partial_n G_{rb} + \kappa G_{rb} = 0,\quad\text{on }\partial\Omega,
\end{equation*}
while the second has a membrane-bound source $\xi\in\partial\Omega$
and solves
\begin{equation*}
\Delta G_{rm} - \mu^2 G_{rm} = 0,\quad\text{in }\Omega,\qquad \partial_n G_{rm} + \kappa G_{rm} = \delta_{\partial\Omega}(x-\xi),\quad\text{on }\partial\Omega,
\end{equation*}
where
\begin{equation*}
\mu = \frac{1}{\sqrt{D_w}},\qquad \kappa = \frac{K_2(1-K_1)}{D_w}.
\end{equation*}
Note that since $K_1<1$ we have $\kappa>0$ and so the problems for the
Robin Green's functions are well-posed. An explicit formula is
available for the membrane Green's function while the Robin Green's
functions are given in terms of series of special functions. We refer
the reader to Appendix \ref{app:green} for relevant formulas and
properties of these Green's functions. When $|x-x_i|=O(1)$ for all
$i=1,...,N$ we therefore have
\begin{equation*}
w(x) = \frac{E_0}{D_w}G_{rb}(x,x_0) + \frac{f}{1-f}\frac{2\pi K_1 D_v^{1/2}}{D_w} \sum_{i=1}^N S_i G_{rm}(x,x_i).
\end{equation*}
Integrating the membrane-bound equation \eqref{eq:pde-outer-v} over
$\partial\Omega$ yields the solvability condition
\begin{equation}\label{eq:pde-solvability}
\biggl(1 - \frac{f}{1-f}\frac{K_1K_2}{D_w}g_0(1)\biggr) \sum_{i=1}^N S_i = \frac{E_0 K_2}{2\pi D_w D_v^{1/2}}g_0(\eta_0),
\end{equation}
where we have used
$$
\int_{\partial\Omega} G_{rm}(x,x_i) dA_x = g_0(1),\qquad \int_{\partial\Omega} G_{rb}(x,x_0)dA_x = g_0(\eta_0),
$$
and each of $g_0(1)$ and $g_0(\eta_0)$ are given explicitly in
\eqref{eq:series-gl}. Since each $S_1,...,S_N>0$ it follows that spots
can be constructed only if $K_1 < K_1^\star$ where
\begin{equation*}
K_1^\star(K_2,D_w,f) = \begin{cases} 1, & K_2 \leq K_2^\star(D_w,f),\\ (1-f)\bigl(1 + \frac{f}{1-f}\frac{K_2^\star}{K_2}\bigr), & K_2 > K_2^\star(D_w,f),\end{cases}
\end{equation*}
and
\begin{equation*}
K_2^\star(D_w,f) = \frac{1-f}{f}\frac{1}{D_w^{-1/2}}\frac{I_{3/2}(D_w^{-1/2})}{I_{1/2}(D_w^{-1/2})}.
\end{equation*}
This existence constraint is a direct consequence of the recirculation
of the membrane bound activator. If the coupling constant $K_1$ is too
high, recirculation is too strong and the feedback loop can no longer
sustain spots with our scaling. The parameter dependence of the
existence threshold is illustrated in Figure
\ref{fig:existence-threshold-illustration}.

If the solvability condition \eqref{eq:pde-solvability} is satisfied
then the solution to \eqref{eq:pde-outer-v} is given by
\begin{equation}
v(x) = -\frac{2\pi}{\sqrt{D_v}}\sum_{j=1}^N S_j G_m(x,x_j) + \frac{K_2}{D_v}v_{1p}(x) + \frac{\bar{v}}{\sqrt{D_v}},
\end{equation}
where $\bar{v}$ is an undetermined constant and $v_{1p}$ is the unique
solution to
\begin{equation}
\Delta_{\partial\Omega}v_{1p} = \frac{1}{|\partial\Omega|}\int_{\partial\Omega}w dA - w\quad\text{in }\partial\Omega;\qquad \int_{\partial\Omega} v_{1p}(x)dA = 0,
\end{equation}
given by
\begin{equation}
v_{1p}(x) = \int_{\partial\Omega} G_m(x,\xi) w(\xi)dA_\xi.
\end{equation}
We match this to the inner solution by calculating expansion of $v(x)$
as $x$ approaches each spot. Using \eqref{eq:green-membrane-exact}
with \eqref{eq:local-global-norm-x-x_i} and
\eqref{eq:local-global-norm-x-x_i-xi} we calculate that for
$|x-x_i|=O(\varepsilon)$ but $|\xi-x_i|=O(1)$
\begin{equation*}
G_m(x,x_i)\sim -\frac{1}{2\pi}\log\rho + R + \frac{1}{2\pi\nu} + O(\varepsilon^2),\quad G_m(x,\xi) \sim G_m(x_i,\xi) + \frac{1}{2\pi}\frac{y^T \mathcal{J}_i^T(\xi-x_i)}{|\xi-x_i|^2}\varepsilon + O(\varepsilon^2),
\end{equation*}
where
\begin{equation}\label{eq:nu-def}
\nu := -\frac{1}{\log\varepsilon}.
\end{equation}
Therefore as $|x-x_i|\rightarrow 0$ we have
\begin{align*}
& \sum_{j=1}^N S_j G_m(x,x_j) \sim S_i\biggl(-\frac{1}{2\pi}\log\rho + \frac{1}{2\pi\nu} + R \biggr) + \sum_{j\neq i} S_j \biggl( G_m(x_i,x_j) + \frac{\varepsilon}{2\pi} \frac{y^T \mathcal{J}_i^T(x_j-x_i)}{|x_j-x_i|^2}  \biggr) + o(\varepsilon), \\
& \int_{\partial\Omega} G_m(x,\xi) G_{rb}(\xi,x_0) dA_\xi \sim \int_{\partial\Omega} G_m(x_i,\xi)G_{rb}(\xi,x_0)dA_\xi + \frac{\varepsilon}{2\pi}y^T \mathcal{J}_i^T \int_{\partial\Omega}\frac{\xi-x_i}{|\xi-x_i|^2} G_{rb}(\xi,x_0)dA_\xi + o(\varepsilon), \\
& \int_{\partial\Omega} G_m(x,\xi) G_{rm}(\xi,x_j) dA_\xi \sim \int_{\partial\Omega} G_m(x_i,\xi)G_{rm}(\xi,x_j)dA_\xi + \frac{\varepsilon}{2\pi}y^T \mathcal{J}_i^T \int_{\partial\Omega}\frac{\xi-x_i}{|\xi-x_i|^2} G_{rm}(\xi,x_j)dA_\xi + o(\varepsilon).
\end{align*}
The behaviour of the outer solution $v(x)$  as $|x-x_i| \rightarrow 0$ is thus given by
\begin{equation}\label{eq:v-outer-limit}
\begin{split}
v(x) \sim & \frac{1}{\sqrt{D_v}}\biggl[ S_i\biggl(\log\rho -2\pi R - \frac{1}{\nu}\biggr) -2\pi \sum_{j\neq i} S_j G_m(x_i,x_j)  + \bar{v} +  \frac{E_0K_2}{D_w\sqrt{D_v}}\int_{\partial\Omega} G_m(x_i,\xi)G_{rb}(\xi,x_0)dA_\xi \\
&  + 2\pi \frac{f}{1-f}\frac{K_1K_2}{D_w}\sum_{j=1}^{N}S_j\int_{\partial\Omega} G_m(x_i,\xi)G_{rm}(\xi,x_j)dA_\xi \biggr] +  \frac{\varepsilon}{\sqrt{D_v}}y^T \mathcal{J}_i^T \biggl[ -\sum_{j\neq i}S_j\frac{x_j - x_i}{|x_j-x_i|^2} \\
&  + \frac{E_0K_2}{2\pi D_w \sqrt{D_v}} \int_{\partial\Omega}\frac{\xi-x_i}{|\xi-x_i|^2} G_{rb}(\xi,x_0)dA_\xi + \frac{f}{1-f}\frac{K_1K_2}{D_w} \sum_{j\neq i}S_j \int_{\partial\Omega}\frac{\xi-x_i}{|\xi-x_i|^2} G_{rm}(\xi,x_j)dA_\xi\biggr] + o(\varepsilon),
\end{split}
\end{equation}
where the $j=i$ term in the last sum vanishes due to rotational
symmetry. Equating the $O(1)$ term to the limiting behaviour of
$D_v^{-1/2}v_{i0}(\rho)$ as $\rho\rightarrow\infty$ given in
\eqref{eq:core-problem-membrane} yields the matching equation
\begin{equation}\label{eq:matching-components}
\begin{split}
(1 + 2 \pi \nu R) S_i &  + 2\pi \ \nu \sum_{j\neq i}S_j G_m(x_i,x_j) - 2\pi\nu\frac{f}{1-f}\frac{K_1K_2}{D_w}\sum_{j=1}^N S_j\int_{\partial\Omega} G_m(x_i,\xi)G_{rm}(\xi,x_j) dA_\xi + \nu\chi(S_i,f) \\
& = \nu\frac{E_0 K_2}{D_w \sqrt{D_v}} \int_{\partial\Omega} G_m(x_i,\xi)G_{rb}(\xi,x_0)dA_\xi + \nu \bar{v}.
\end{split}
\end{equation}
We write this in a more convenient way by first defining the Green's
matrix
\begin{equation}
\mathcal{G} := \mathcal{G}_m - \frac{f}{1-f}\frac{K_1K_2}{D_w} \mathcal{G}_{rm},
\end{equation}
where the matrices $\mathcal{G}_m$ and $\mathcal{G}_{rm}$ have entries
\begin{equation}
(\mathcal{G}_m)_{ij} := \begin{cases} R & i=j \\ G_m(x_i,x_j) & i\neq j \end{cases},\qquad (\mathcal{G}_{rm})_{ij} := \int_{\partial\Omega} G_m(x_i,\xi)G_{rm}(\xi,x_j) dA_\xi.
\end{equation}
We also define the vectors $\pmb{e}$, $\pmb{\chi}$, and $\pmb{g}_{rb}$ by
\begin{equation}
\pmb{e} = \begin{pmatrix} 1 \\ \vdots \\ 1 \end{pmatrix},\qquad \pmb{S} = \begin{pmatrix} S_1 \\ \vdots \\ S_N \end{pmatrix},\qquad \pmb{\chi}(\pmb{S}) = \begin{pmatrix} \chi(S_1,f) \\ \vdots \\ \chi(S_N,f) \end{pmatrix},\qquad \pmb{g}_r = \begin{pmatrix} \int_{\partial\Omega} G_m(x_1,\xi)G_{rb}(\xi,x_0)dA_\xi \\ \vdots \\ \int_{\partial\Omega} G_m(x_N,\xi)G_{rb}(\xi,x_0)dA_\xi \end{pmatrix}.
\end{equation}
With these definitions \eqref{eq:pde-solvability} and \eqref{eq:matching-components} become
\begin{align}
\pmb{e}^T \pmb{S} = N S_c, \qquad \bigl(\mathbb{I}_N + 2\pi\nu\mathcal{G}\bigr)\pmb{S} + \nu \pmb{\chi}(\pmb{S}) = \frac{\nu E_0 K_2}{D_w D_v^{1/2}} \pmb{g}_{rb} + \nu \bar{v}\pmb{e},
\end{align}
where
\begin{equation}
S_c = \frac{1}{N}\frac{\frac{E_0K_2}{2\pi D_w \sqrt{D_v}} g_0(\eta_0)}{1 - \frac{f}{1-f}\frac{K_1K_2}{D_w}g_0(1)}.
\end{equation}
Left multiplying the second equation by $\pmb{e}^T$ and substituting
the first we get
\begin{equation}
\bar{v} = \frac{S_c}{\nu} + \frac{1}{N}\biggl( 2\pi\pmb{e}^T \mathcal{G} \pmb{S} + \pmb{e}^T\pmb{\chi} - \frac{E_0K_2}{D_w\sqrt{D_v}}\pmb{e}^T\pmb{g}_{rb} \biggr).
\end{equation}
The spot strengths $S_1,...,S_N$ are therefore determined
by solving the Nonlinear Algebraic System (NAS)
\begin{equation}\label{eq:NAS}
\pmb{S} + 2\pi\nu(\mathbb{I}_N - \mathcal{E}_N)\mathcal{G} \pmb{S} + \nu (\mathbb{I}_N - \mathcal{E}_N)\pmb{\chi} = S_c \pmb{e} + \frac{\nu E_0K_2}{D_w \sqrt{D_v}}(\mathbb{I}-\mathcal{E}_N)\pmb{g}_{rb},
\end{equation}
where
\begin{equation}\label{eq:definition-E}
\mathcal{E}_N := \frac{1}{N}\pmb{e}\pmb{e}^T.
\end{equation}

In the absence of coupling, the NAS \eqref{eq:NAS} is known to have a
rich bifurcation structure \cite{trinh2016}. By maintaining that our
parameters are $O(1)$ with respect to $\varepsilon$ we restrict
ourselves to $O(1)$ spot patterns. This simplification allows us to
more clearly explore the effects of coupling on stability and dynamics
of $O(1)$ spot patterns. An important case for us is when the points
$x_1,...,x_N$ are uniformly distributed on a ring making a common
angle with $x_0$. Then it is easy to see that $\mathcal{G}$ has
constant row sum and $\pmb{g}_{rb}$ is proportional to $\pmb{e}$. In
such a case $\pmb{S} = S_c\pmb{e}$ is an exact solution to
\eqref{eq:NAS}. We highlight here that as $K_1$ approaches the
existence threshold $K_1^\star$ the common spot strength $S_c$ grows
to infinity. This suggests that the scaling used to derive
\eqref{eq:pde-all} is no longer valid in this parameter and we leave
the analysis of alternative scalings for future studies.

\section{Linear Stability: $O(1)$ Eigenvalues}\label{sec:stability}

The linear stability of the quasi-equilibrium solution constructed
above is determined by a non-linear eigenvalue problem which we derive
below. In our analysis we make two simplifying assumptions. First, we
assume that $\lambda = O(1)$ with respect to the small parameter
$\varepsilon$. The remaining small eigenvalues lead to drift
instabilities and their effect is described by the slow dynamics ODE
analysed in the next section. Second, we will focus only on
instabilities caused by a zero eigenvalue crossing and therefore neglect the possibility of H\"opf bifurcations. In previous studies of the uncoupled Brusselator model it has been shown that the existence of H\"opf instabilities is closely related to the choice of time constant $\tau_v>0$ \cite{rozada2014}. By appropriately choosing values of the parameters $\tau_v$ and $\tau_w$ in the proceeding sections we will therefore assume that H\"opf instabilities are avoided.

Linearizing about the quasi-equilibrium solution
$$
u = u_e + e^{\lambda t}\hat{\phi},\quad v = v_e + e^{\lambda t}\hat{\psi},\quad w = w_e + e^{\lambda t}\hat{\eta},
$$
we obtain the eigenvalue problem
\begin{subequations}
\begin{align}
&\varepsilon^2 \Delta_{\partial\Omega} \hat{\phi} - \hat{\phi} + \varepsilon^2 K_2 \hat{\eta} + 2fu_ev_e \hat{\phi} + fu_e^2\hat{\psi} = \lambda \hat{\phi}, & \text{in }\partial\Omega,\\
& D_v\Delta_{\partial\Omega}\hat{\psi} + \varepsilon^{-2}[\hat{\phi} - 2u_e v_e \hat{\phi} - u_e^2 \hat{\psi}] = \tau_v\lambda\hat{\psi}, & \text{in }\partial\Omega, \\
& D_b\Delta\hat{\eta} - \hat{\eta} = \tau_w\lambda\hat{\eta}, & \text{in }\Omega,\\
& D_b\partial_n \hat{\eta} + K_2 \hat{\eta} = \varepsilon^{-2}K_1\hat{\phi}, & \text{on }\partial\Omega.
\end{align}
\end{subequations}
We can reduce this to an algebraic system by once again using the
method of matched asymptotic expansions. We begin by noting that
$\hat{\phi}$ is strongly localized, while $\hat{\eta}$ is bounded and
$O(\varepsilon)$ near each $x_j$. Near each $x_j$ we use the local
coordinates \eqref{eq:local-coordinates} and introduce the inner
solution
$$
\hat{\phi} \sim \phi_j(\rho) e^{im\omega},\qquad \hat{\psi} \sim D_v^{-1}\psi_j(\rho)e^{im\omega},
$$
where the polar coordinates $(\rho,\omega)$ are defined by
$y_1 = \rho\cos\omega$ and $y_2=\rho\sin\omega$. Assuming that
$\tau_v\lambda\varepsilon^2\ll 1$ the membrane and bulk problems
decouple leading to the inner eigenvalue problems
\begin{equation}\label{eq:eigenvalue-problem-inner}
\Delta_\rho \Psi_j - \frac{m^2}{\rho^2}\Psi_j + \mathcal{Q}_j \Psi_j = \lambda \mathcal{E}_{11}\Psi_j\qquad\text{for each } j = 1,...,N,
\end{equation}
where
\begin{equation}
\mathcal{Q}_j := \begin{pmatrix}2fu_{j0}v_{j0}-1 & fu_{j0}^2 \\ -2u_{j0}v_{j0}+1 & -u_{j0}^2 \end{pmatrix},\quad \mathcal{E}_{11} = \begin{pmatrix} 1 & 0 \\ 0 & 0 \end{pmatrix},\quad \Psi_j := \begin{pmatrix}\phi_j \\ \psi_j\end{pmatrix}.
\end{equation}
Smoothness of the eigenfunctions imposes the boundary condition
$\Psi_{j}'(0) = 0$ while the behaviour as $\rho\rightarrow\infty$ is
determined by noting that
$$
\mathcal{Q}_j\sim \begin{pmatrix} -1 & 0 \\ 1 & 0\end{pmatrix}\qquad\text{as }\rho\rightarrow\infty,
$$
leading to the limiting system
$$
\phi_j'' + \frac{1}{\rho}\phi_j' - \frac{m^2}{\rho^2}\phi_j - (1+\lambda)\phi_j \sim 0,\qquad \psi_j'' + \frac{1}{\rho}\psi_j' - \frac{m^2}{\rho^2}\psi_j + \phi_j \sim 0,\qquad\text{as }\rho\rightarrow\infty.
$$
Since we are looking for conditions under which an eigenvalue crosses
the imaginary axis, we may assume that $\text{Re}\lambda > -1$ and
therefore
\begin{equation}\label{eq:eigenfunction-decay}
\phi_j \rightarrow 0\quad\text{and}\quad \psi_j\sim\begin{cases} O(\log\rho) & m=0 \\ O(\rho^{-m}) & m\geq 1,\end{cases}\qquad\qquad\text{as }\rho\rightarrow\infty.
\end{equation}
Since $m=1$ corresponds to the translational mode with neutral
eigenvalue $\lambda = 0$, the linear stability is determined by the
$m=0$ and $m\geq 2$ modes. We study these two cases separately since
the logarithmic growth of $\psi_j$ for $m=0$ and its algebraic decay
for $m\geq 2$ lead to two very different instability mechanisms.

\subsection{The $m\geq 2$ Mode Instabilities}

The algebraic decay of $\psi_j(\rho)$ as $\rho\rightarrow\infty$ when
$m\geq 2$ leads to a decoupling of the inner eigenvalue problems
\eqref{eq:eigenvalue-problem-inner}, with the only global coupling
arising through the NAS \eqref{eq:NAS} relating the source strengths
$S_1,...,S_N$. In this sense we deduce that the $m\geq 2$ modes lead
to strongly local instabilities. Note that this decoupling implies
that the $m\geq 2$ instabilities are identical to those
found in the studies of \cite{rozada2014,trinh2016}.

It is easy to see that for each $j=1,..,N$ the corresponding
eigenvalue depends only on $S_j$. Omitting subscripts, we can
therefore calculate $\lambda$ as a function of $S$ by simultaneously
solving \eqref{eq:core-problem-membrane} for $(u_0,v_0)$ and then
calculating the eigenvalue with largest real part of the problem
\eqref{eq:eigenvalue-problem-inner}. This is performed numerically by
discretizing the eigenvalue problem
\eqref{eq:eigenvalue-problem-inner} on a truncated domain
$0 < \rho < L$ and computing the eigenvalues of the resulting
matrix. We outline here only the main results and refer to \S 3.1 of
\cite{rozada2014} for more details. The outcome of these computations
is that the dominant eigenvalue of \eqref{eq:eigenvalue-problem-inner}
crosses to the unstable half-plane through a zero-eigenvalue crossing
when $S$ exceeds a threshold $\Sigma_m(f)$. In Figure
\ref{fig:splitting_instability} we plot these thresholds as functions
of $f$ for fixed values of $m$. The ordering
$\Sigma_2(f) < \Sigma_3(f) < \cdots$ indicates that if
$S > \Sigma_2(f)$ the spot will succumb to one of the $m\geq 2$ mode
instabilities.

The stability criteria from this analysis is clear. For a given spot
configuration $\{x_i\}_{i=1}^N$ we compute $S$ by solving the NAS
\eqref{eq:NAS} and label the configuration as unstable with respect to
the $m\geq 2$ mode instabilities if any $S_i$ exceeds the value of
$\Sigma_2(f)$. Previous numerical experiments (see for
examples \cite{rozada2014}) reveal that the $m=2$ linear spot
shape-deformation instability triggers a nonlinear event resulting
in the formation of two identical spots. In this sense, we will
refer to the $m=2$ mode instability as a ``spot-splitting instability.''  

To highlight the effects of the coupling parameters $K_1$ and $K_2$ on
the splitting instabilities we consider a symmetric $N$-spot pattern
with common spot strength $S_1=...=S_N= S_c$. The splitting
instability threshold is found by setting
$$
S_c = \Sigma_2(f),
$$
from which we can explicitly calculate
\begin{equation}\label{eq:splitting-threshold-common}
K_{1\Sigma} = \min\biggl\{ K_1^\star, (1-f)\biggl(1 + \frac{f}{1-f}\frac{K_2^\star}{K_2} - \frac{\zeta}{2\pi N \sqrt{\xi}}\frac{1}{\Sigma_2}  \biggr)\biggr\},
\end{equation}
where
$$
\xi := \frac{D_v}{E_0^2},\qquad \zeta := \frac{1}{\sqrt{\eta_0}}\frac{I_{1/2}(\mu\eta_0)}{I_{1/2}(\mu)},
$$
and for which splitting instabilities are triggered when $K_1 > K_{1\Sigma}$.

The threshold $S > \Sigma_2(f)$ indicates that a spot succumbs to
splitting instabilities when it is too large. The threshold
\eqref{eq:splitting-threshold-common} concisely indicates
the four ways that this is possible for symmetric spots. First,
increasing $K_1$ results in stronger recirculation therefore
increasing $S_c$ and leading to spot splitting instabilities. Second,
increasing $K_2$ amplifies the recirculation effect and lowers the
threshold $K_{1\Sigma}$. Third, by either reducing the membrane
diffusivity $D_v$ or increasing the bulk-source strength $E_0$, both
of which decrease $\xi$, we promote larger spots and therefore reduce
the splitting threshold $K_{1\Sigma}$. Finally, increasing $\eta_0$
leads to a stronger source term in the membrane and
therefore larger spots. This effect is captured by noting that $\zeta$
is increasing in $\eta_0$.

%\begin{figure}
%	\centering
%	\includegraphics[scale=0.6]{figures/stability/sc_contours_f_40.pdf}
%	\includegraphics[scale=0.6]{figures/stability/sc_contours_f_60.pdf}
%	\caption{Plot of $\bigl(1 + \tfrac{f}{1-f}\tfrac{K_2^*}{K_2} - \tfrac{1}{1-f}K_1\bigr)^{-1}$ for $f=0.4$ and $f=0.6$}\label{fig:Sc_contour}
%\end{figure}

%\begin{figure}
%	\centering
%	\begin{subfigure}[t]{0.5\textwidth}
%		\centering
%		\includegraphics[height=2.5in]{figures/stability/zeta_term.pdf}
%		\caption{}
%	\end{subfigure}%
%	~ 
%	\begin{subfigure}[t]{0.5\textwidth}
%		\centering
%		\includegraphics[height=2.5in]{figures/stability/xi_term.pdf}
%		\caption{}
%	\end{subfigure}
%	\caption{(a) Dependence of $\zeta^{-1}$ on the bulk diffusivity $D_w$ and source location $\eta_0$. (b) Dependence of common soure multiplier on $\xi=D_vE_0^{-2}$ and $f$.}\label{fig:zeta_and_xi}
%\end{figure}

\subsection{The $m=0$ Mode Zero-Eigenvalue Crossing Instabilities}

Seeking instabilities triggered by a zero-eigenvalue crossing we will
henceforth assume $\lambda = 0$. The logarithmic growth of $\psi_j$ in
\eqref{eq:eigenfunction-decay} indicates that the $m=0$ mode
instabilities are globally coupled. We write the far-field condition
for $\psi_j$ explicitly as
$$
\phi_j \sim 0,\qquad \psi_j \sim c_j\log \rho + b_j,\qquad\text{as }\rho\rightarrow\infty,
$$
where $c_j$ is undetermined. From the homogeneity of
\eqref{eq:eigenvalue-problem-inner} we may rescale
$\Psi_j = c_j \tilde{\Psi}_j$ obtaining an identical problem except
for the far-field condition which now takes the form
$$
\tilde{\phi}_j\sim 0,\qquad \tilde{\psi}_j \sim \log \rho + \tilde{b}_j(S_j,f),\qquad\text{as }\rho\rightarrow\infty,
$$
where $\tilde{b}_j(S_j,f)$ may now be computed numerically. Applying
the divergence theorem to $\Delta (\phi_j + f\psi_j)$ and
$\Delta \psi_j$ we obtain the useful identities
\begin{equation}\label{eq:local_integrals}
\int_{\mathbb{R}^2}\phi_j dy = \frac{2\pi f}{1 - f} c_j,\qquad \int_{\mathbb{R}^2}\bigl[(1-2u_{j0}v_{j0})\phi_j - u_{j0}^2\bigr] dy = -2\pi c_j.
\end{equation}
From the exponential decay of $\phi_j$ we obtain the approximation for
the outer solution
$$
\hat{\phi}(x) = \varepsilon^2 K_2 \hat{\eta} + \sum_{j=1}^{N} \phi_j\bigl(\tfrac{|x-x_j|}{\varepsilon}\bigr).
$$
Using \eqref{eq:local_integrals} we then deduce the distributional limits
distributional limits
\begin{equation}\label{eq:lin-stab-dist-lim}
\frac{1}{\varepsilon^2}\hat{\phi} \longrightarrow K_2\hat{\eta} + \frac{2\pi f}{1 - f}\sum_{j=1}^N c_j \delta_{\partial\Omega}(x-x_j),\quad \frac{1}{\varepsilon^2}\bigl[(1 - 2u_ev_e)\hat{\phi} - u_e^2\hat{\psi}\bigr]\longrightarrow K_2\hat{\eta} - 2\pi\sum_{j=1}^Nc_j\delta_{\partial\Omega}(x-x_j),
\end{equation}
as $\varepsilon\rightarrow 0^+$. Using the first of these, the outer
problem for $\hat{\eta}$ therefore becomes
$$
D_w\Delta\hat{\eta} - \hat{\eta} = 0,\quad\text{in }\Omega;\qquad D_w\partial_n\hat{\eta} + K_2(1-K_1)\hat{\eta} = \frac{2\pi f K_1}{1 - f}\sum_{j=1}^Nc_j\delta_{\partial\Omega}(x-x_j),\qquad\text{on }\partial\Omega .
$$
The solution is written in terms of the Green's function $G_{rm}$ as
$$
\hat{\eta} = \frac{2\pi f K_1}{D_w(1 - f)}\sum_{j=1}^N c_j G_{rm} (x,x_j).
$$
Using the second limit in \eqref{eq:lin-stab-dist-lim} we obtain the
outer problem for $\hat{\psi}$
\begin{equation}\label{eq:psi_eigenvalue_prob}
D_v \Delta_{\partial\Omega} \hat{\psi} = -\frac{2\pi K_1K_2 f}{D_w(1 - f)}\sum_{j=1}^N c_j G_{rm} (x,x_j) + 2\pi \sum_{j=1}^N c_j\delta_{\partial\Omega}(x-x_j),\qquad\text{in }\partial\Omega,
\end{equation}
with singular behaviour determined by matching with the inner solution
\begin{equation}\label{eq:psi-inner-limit-behaviour}
\hat{\psi}(x)\sim \frac{c_i}{D_v}\bigl( \log |x-x_i| + \frac{1}{\nu} + \tilde{b}_i \bigr)\qquad\text{as }|x-x_i|\rightarrow 0.
\end{equation}
We are immediately confronted with the solvability condition
$$
\biggl( 1 - \frac{K_1K_2}{D_w}\frac{f}{1 - f}\int_{\partial\Omega} G_{rm}(x,e_z)dA_x \biggr) \sum_{j=1}^N c_j = 0.
$$
Provided $K_1 < K_1^\star$, the solvability condition is satisfied if and
only if
\begin{equation}\label{eq:lin-stab-solvability-0}
\sum_{j=1}^N c_j = 0.
\end{equation}
When this holds, we may solve for $\hat{\psi}$ 
\begin{equation}\label{eq:psi-formula}
\hat{\psi}(x) = -\frac{2\pi}{D_v}\sum_{j=1}^Nc_j G_m(x,x_j) + \frac{1}{D_v}\bar{\psi} + \frac{2\pi K_1K_2 f}{D_wD_v(1 - f )}\sum_{j=1}^N c_j\hat{\psi}_p(x,x_j),
\end{equation}
where $\hat{\psi}_p(x,\xi)$ is the unique solution to
$$
\Delta \hat{\psi}_p  = \frac{1}{|\partial\Omega|}\int_{\partial\Omega} G_{rm}(x,\xi) dA_x - G_{rm}(x,\xi),\quad\text{on }\partial\Omega,\qquad \int_{\partial\Omega}\hat{\psi}_p = 0,
$$
given by
$$
\hat{\psi}_p(x,\xi) = \int_{\partial\Omega} G_{m}(x,y) G_{rm}(y,\xi) dA_y.
$$
Expanding \eqref{eq:psi-formula} as $|x-x_i|\rightarrow 0$ and
comparing to \eqref{eq:psi-inner-limit-behaviour} yields the system
$$
\biggl(\mathbb{I}  + 2\pi\nu \mathcal{G}_m + \nu\tilde{\mathcal{B}} - \frac{2\pi\nu K_1K_2 f}{D_w(1- f)}\mathcal{G}_{rm}\biggr)\pmb{c} = \nu \bar{\psi}\pmb{e},
$$
where
\begin{equation}
\pmb{c} = (c_1,...,c_N)^T,\qquad \tilde{\mathcal{B}} = \text{diag}(\tilde{b}_1(S_1,,f),...,\tilde{b}_N(S_N,f).
\end{equation}
Left multiplying by $\pmb{e}^T$ and using
\eqref{eq:lin-stab-solvability-0} we can isolate for $\bar{\psi}$ and
thus obtain the reduced system in the unknown $\pmb{c}$ given by
$$
\mathcal{M}\pmb{c} = 0.
$$
where
\begin{equation}\label{eq:GCEP}
\mathcal{M}(\pmb{S},f) := \frac{1}{\nu}\mathbb{I}_N + 2\pi\bigl(\mathbb{I}_N - \mathcal{E}_N\bigr)\biggl( \mathcal{G}_m - \frac{K_1K_2 f}{D_w(1- f)}\mathcal{G}_{rm}  \biggr) + \bigl(\mathbb{I}_N - \mathcal{E}_N\bigr)\tilde{\mathcal{B}}.
\end{equation}
Since we are seeking \textit{non-trivial} solutions to this
homogeneous system the instability threshold in parameter space is
found by solving $\det\mathcal{M} = 0$.

The key identity
\begin{equation}
\tilde{b}_j(S_j,f) = \chi'(S_j;f),
\end{equation}
leads to the simplification
\begin{equation}\label{eq:GCEP0}
\mathcal{M}(\pmb{S},f) =\frac{1}{\nu}\mathbb{I}_N + 2\pi\bigl(\mathbb{I}_N - \mathcal{E}_N\bigr)\mathcal{G} + \bigl(\mathbb{I}_N - \mathcal{E}_N\bigr)\text{diag}(\chi'(S_1;f),...,\chi'(S_N;f)),
\end{equation}
which we recognize as the derivative of the NAS \eqref{eq:NAS} with
respect to $\pmb{S}$. Instabilities of the $m=0$ mode that arise
through a zero-eigenvalue crossing therefore correspond to loss of
uniqueness of solutions to the NAS. This observation has been made in
previous studies of the Brusselator on the sphere
\cite{rozada2014,trinh2016} and appears to be a common feature for a
class of singularly perturbed reaction diffusion systems.

We conclude this section by considering spot configurations
$\{x_i\}_{i=1}^N$ satisfying two assumptions. First, we assume the
arrangement is chosen in such a way that the matrix $\mathcal{G}$ has
constant row sum. Second, we suppose that the common source solution
$\pmb{S} = S_c \pmb{e}$ solves the NAS \eqref{eq:NAS}. This is the
case if, for example, the spots are arranged on a ring making a common
angle with the bulk-source $x_0$. Since the matrix $\mathcal{G}$ is
symmetric, we find that its spectrum has the following properties
\begin{equation}
\mathcal{G}\pmb{q}_j = k_j \pmb{q}_j,\qquad\text{with}\quad \pmb{q}_1 = \pmb{e}\quad\text{and}\quad \pmb{q}_j^T\pmb{q}_1 = 0\quad\text{for }\quad j=2,...,N.
\end{equation}
Recalling the definition
$\mathcal{E}_N := N^{-1}\pmb{e}\pmb{e}^T$ we find that the spectrum of
$\mathcal{M}_c := \mathcal{M}(S_c\pmb{e},f)$ is given by
\begin{equation}
\mathcal{M}_c\pmb{q}_1 = \nu^{-1},\qquad \mathcal{M}_c\pmb{q}_j = \mathcal{A}_j \pmb{q}_j\qquad (j\geq 2),
\end{equation}
where
$$
\mathcal{A}_j := \frac{1}{\nu} + 2\pi k_j + \chi'(S_c;f).
$$
The small $S$ asymptotics $\chi(S) \sim \frac{d_0}{S} + d_1 S + o(S)$, where (see equation (4.20) in \cite{rozada2014})
\begin{equation}
d_0 = \frac{b(1-f)}{f^2},\quad d_1 = \frac{0.4893}{1-f} - 0.4698,\qquad b:=\int_0^\infty w^2\rho d\rho \approx 4.934,
\end{equation}
suggest that $S_c = O(\nu^{1/2})$ at the instability
threshold. Furthermore, we expect the $m=0$ mode instability to
persist as $S_c$ is decreased and since $\chi'(S;f)$ is monotone
increasing in $S$ the threshold is determined by the smallest of $k_j$
($j\geq 2$). Thus, the $m=0$ mode instability is determined by solving
\begin{equation}\label{eq:competition-A}
\mathcal{A}_c^\star = 0,\qquad \mathcal{A}_c^\star := \frac{1}{\nu} + 2\pi k^\star + \chi'(S_c;f),\quad k^\star :=\min_{2\leq j\leq N} k_j.
\end{equation}
We remark that the eigenvector $\pmb{q}^\star$ corresponding to
$k^\star$ will satisfy $\pmb{e}^T\pmb{q}^\star = 0$. Thus when an
$m=0$ mode instability is triggered it will cause a net-zero increase
in the heights of individual spots. For this reason, $m=0$ mode
instabilities are typically referred to as ``competition
instabilities.''

To determine the competition instability threshold we
therefore have to solve \eqref{eq:competition-A}
numerically. The leading order balance between $\nu^{-1}$ and
$\chi'(S_c,f)$ yields an approximate value of $S_c\sim S_{c0}^\star$
independent of all problem parameters except $f$ and determined by
solving
$$
\frac{1}{\nu}  + \chi'(S_{c0}^\star; f) = 0.
$$
From this we easily determine a leading order approximation for the instability threshold in $(K_1,K_2)$ parameter space
\begin{equation}\label{eq:competition-leading-order}
K_{1C0} \sim \min\biggl\{ K_1^\star, (1-f)\biggl(1 + \frac{f}{1-f}\frac{K_2^\star}{K_2} - \frac{\zeta}{2\pi N \sqrt{\xi}}\frac{1}{S_{c0}^\star}  \biggr)\biggr\},
\end{equation}
which can be used to facilitate the numerical solution of
$\mathcal{A}_c^\star = 0$. Note that since the competition
instabilities persist as $S_c$ is decreased we deduce that competition
instabilities are triggered whenever $K_1 < K_{1C}$.

\begin{figure}[t!]
	\centering
	\begin{subfigure}{0.5\textwidth}
		\centering
		\includegraphics[scale=0.75]{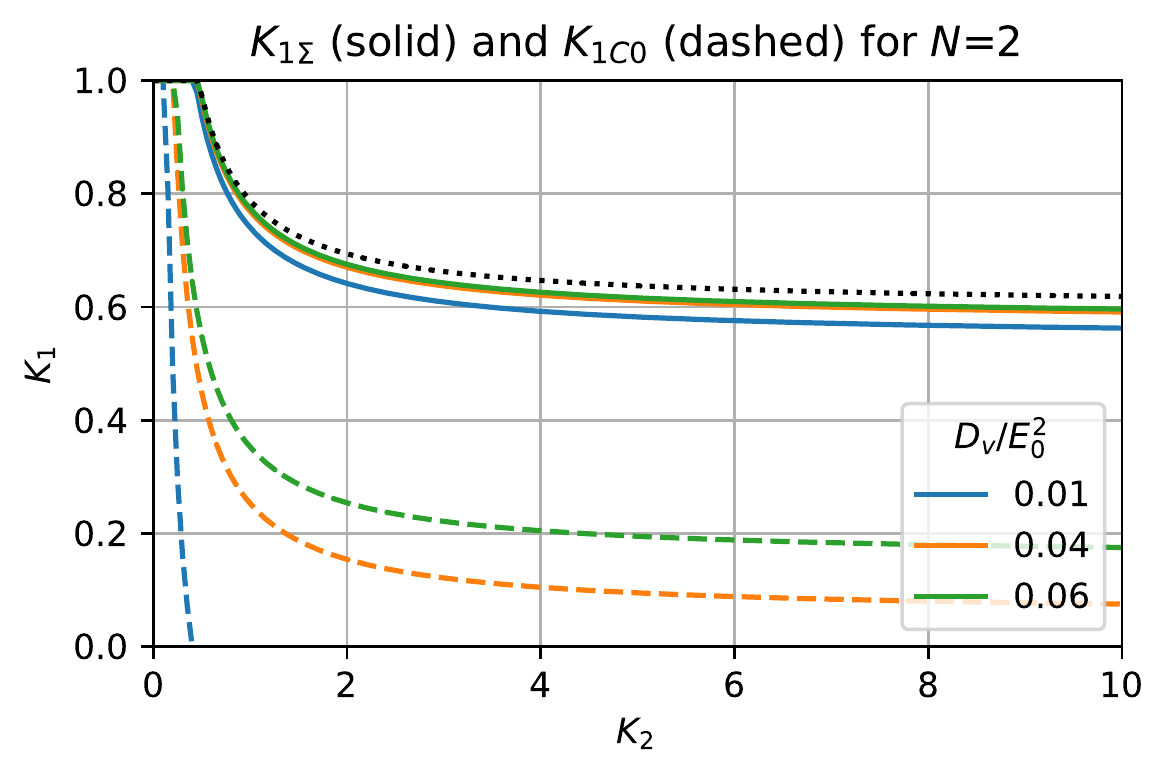}
		\caption{}\label{fig:comp_and_split_2}
	\end{subfigure}%
	\begin{subfigure}{0.5\textwidth}
		\centering
		\includegraphics[scale=0.75]{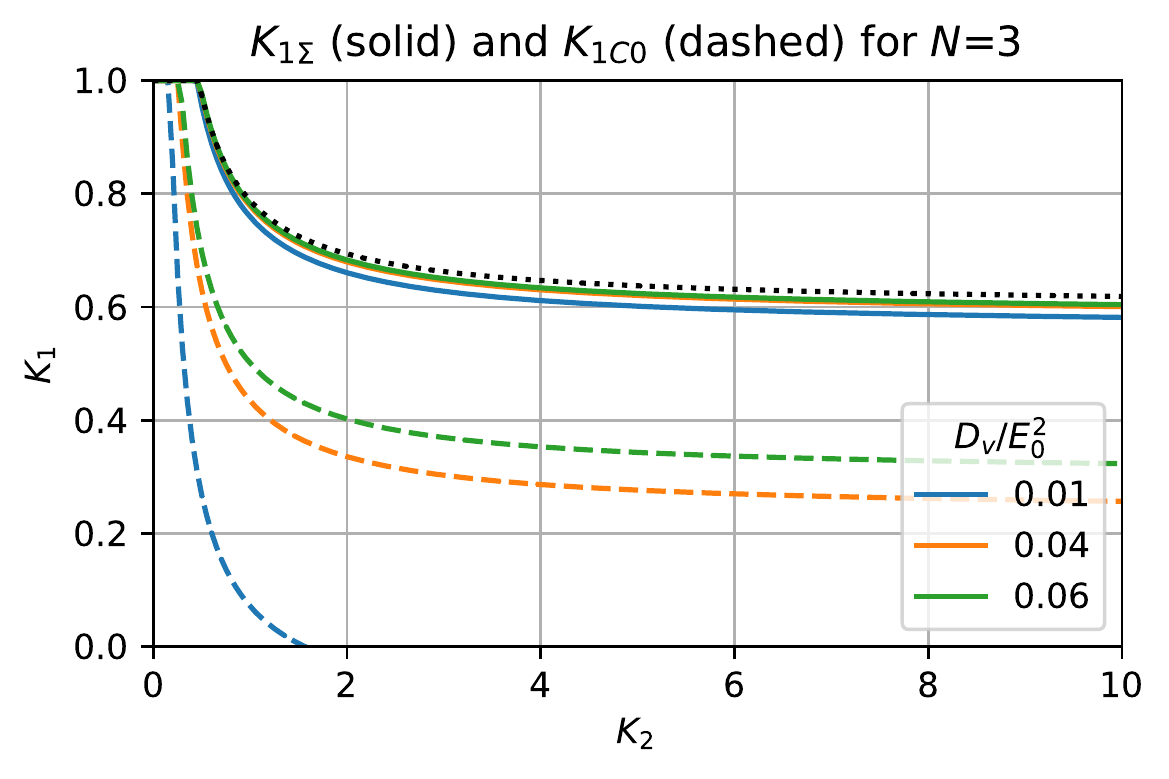}
		\caption{}\label{fig:comp_and_split_3}
	\end{subfigure}%
	\caption{Plots of the leading order competition instability threshold $K_{1C0}$ (dashed lines) and the splitting instability threshold $K_{1\Sigma}$ (solid lines) for (a) two-spot configuration, and (b) a three-spot configuration. In both Figures the legend applies to both the dashed and solid lines. Fixed parameters are $f = 0.4$, $\eta_0=0.6$, $D_w=5$, and $\nu=0.01$.}\label{fig:splitting-competition-plot}
\end{figure}

We conclude this section by noting that since
$S_{c0}^\star = O(\nu^{1/2})$ while $\Sigma_2(f)=O(1)$ we have that
$K_{1C0} \leq K_{1\Sigma}$. Figure
\ref{fig:splitting-competition-plot} illustrates the general behaviour of these instability thresholds as parameter are varied. Note in particular that increasing the number of spots will expand the stability region for splitting instabilities but reduce it for competition instabilities.

\section{Slow-Spot Dynamics}\label{sec:dynamics}

In this section we derive an ODE system that describes the slow-time
evolution of an $N$-spot configuration. Specifically we assume that
$x_i = x_i(\sigma)$ where $\sigma$ is a slow time variable which is
determined by a dominant balance to be $\sigma=\varepsilon^2 t$. Next
we differentiate \eqref{eq:x-y-relationship-1} with respect to
$\sigma$ and use the identities \eqref{eq:J_i_identities} to deduce
that for the local coordinates $y$ near the $i^{\text{th}}$ spot
defined by \eqref{eq:local-coordinates} we have
\begin{equation}
\frac{d y}{d \sigma} = -\varepsilon^{-1}\mathcal{T}_i + O(1),\qquad\text{where}\quad \mathcal{T}_i = \begin{pmatrix} \sin\theta_i\frac{d\varphi_i}{d\sigma} \\ \frac{d\theta_i}{d\sigma}\end{pmatrix},
\end{equation}
and therefore
\begin{equation}
\partial_t = -\varepsilon \mathcal{T}_i\cdot\nabla_y + O(\varepsilon^2),\qquad \text{where} \quad \nabla_y = \begin{pmatrix} \partial_{y_1} \\ \partial_{y_2} \end{pmatrix}.
\end{equation}
The subsequent analysis follows closely that found in
\cite{trinh2016}. The key idea is to perform a higher order asymptotic
matching from which a solvability condition yields an equation for
$\mathcal{T}_i$.

Near the $i^\text{th}$ spot we consider the two term expansions
\begin{equation*}
u \sim D_v^{1/2} ( u_{i0}(\rho) + \varepsilon u_{i1}(\sigma,y)) + o(\varepsilon),\quad v \sim D_v^{-1/2}( v_{i0}(\rho) + \varepsilon v_{i1}(\sigma,y) ) + o(\varepsilon),
\end{equation*}
and leading order expansion
$w \sim \frac{K_1 D_v^{1/2}}{\varepsilon D_w} w_{i0}(y,y_3) +
o(\varepsilon^{-1})$ where the leading order terms are those found in
\S\ref{sec:quasi-equilibrium}. Substituting into \eqref{eq:pde-all}
and collecting powers of $\varepsilon$ results in the linear system
\begin{equation}\label{eq:q_i1_system}
\Delta_y \pmb{q}_{i1}(y) + \mathcal{Q}_{i} \pmb{q}_{i1}(y) =  - \pmb{f}_i,\qquad y\in\mathbb{R}^2,
\end{equation}
where
\begin{equation*}
\pmb{q}_{i1}(y) := \begin{pmatrix}u_{i1}(y) \\ v_{i1}(y) \end{pmatrix},\quad \mathcal{Q}_i := \begin{pmatrix} -1 + 2 fu_{i0}v_{i0} & fu_{i0}^2 \\ 1 - 2u_{i0}v_{i0} & -u_{i0}^2 \end{pmatrix},\quad \pmb{f}_i := \begin{pmatrix} \mathcal{T}_i\cdot\nabla_y u_{i0} + \frac{K_1K_2}{D_w}w_{i0}|_{y_3=0} \\ 0 \end{pmatrix}.
\end{equation*}
The decay of $u_{i0}$ and $w_{i0}$ implies that $u_{i1}\rightarrow 0$
as $\rho\rightarrow\infty$. The limiting behaviour of $v_{i1}$ as
$\rho\rightarrow\infty$ is determined by matching to the limiting
behaviour \eqref{eq:v-outer-limit} of the outer solution as
$|x-x_i|\rightarrow 0$.  Thus
\begin{equation*}
\pmb{q}_{i1}(y) \sim \begin{pmatrix} 0 \\ y^T \mathcal{J}_i^T \pmb{\alpha}_i \end{pmatrix}\qquad\text{as}\quad \rho\rightarrow\infty,
\end{equation*}
where{\small
\begin{equation}
\pmb{\alpha}_i =  \sum_{j\neq i} S_j \biggl( \frac{x_i - x_j}{|x_i-x_j|^2} + \frac{f}{1-f}\frac{K_1K_2}{D_w}\int_{\partial\Omega}\frac{\xi-x_i}{|\xi-x_i|^2} G_{rm}(\xi,x_j)dA_\xi\biggr) + \frac{E_0 K_2}{2\pi D_w\sqrt{D_v}}\int_{\partial\Omega}\frac{\xi-x_i}{|\xi-x_i|^2} G_{rb}(\xi,x_0)dA_\xi.
\end{equation}}

To solve \eqref{eq:q_i1_system} we must first impose a solvability
condition on $\pmb{f}_i$. Indeed, by differentiating
\eqref{eq:core-problem-membrane} with respect to $y_1$ or $y_2$ we
deduce that $\Delta_y + \mathcal{Q}_i$ has a null space of dimension
at least two. We assume that this null space is exactly
two-dimensional and write solutions of the adjoint problem
\begin{equation*}
\Delta_y \pmb{\Psi}(y) + \mathcal{Q}_{i}^T \pmb{\Psi}(y) =  0,\qquad y\in\mathbb{R}^2;\qquad \pmb{\Psi}\rightarrow \begin{pmatrix} 0 \\ 0 \end{pmatrix},
\end{equation*}
in terms of the polar coordinates $(\rho,\omega)$ defined by $y_1 = \rho\cos\omega$ and $y_2 = \rho\sin\omega$ as
\begin{equation*}
\pmb{\Psi}_c = \pmb{P}(\rho)\cos\omega,\qquad \pmb{\Psi}_s = \pmb{P}(\rho)\sin\omega,\qquad\text{where}\quad \pmb{P}(\rho) = \begin{pmatrix} P_1(\rho) \\ P_2(\rho) \end{pmatrix}.
\end{equation*}
It follows that $\pmb{P}$ satisfies
\begin{equation}
\pmb{P}''(\rho) + \frac{1}{\rho}\pmb{P}'(\rho) - \frac{1}{\rho^2}\pmb{P}(\rho) + \mathcal{Q}_i^T\pmb{P}(\rho) = 0,\quad\text{in }\rho>0;\qquad \pmb{P}\sim \begin{pmatrix} \rho^{-1} \\ \rho^{-1}\end{pmatrix}\qquad\text{as }\rho\rightarrow\infty,
\end{equation}
where the limiting behaviour as $\rho\rightarrow\infty$ is obtained from
$$
\mathcal{Q}_i^T \rightarrow \begin{pmatrix} -1 & 1 \\ 0 & 0 \end{pmatrix},\qquad\text{as } \rho\rightarrow\infty.
$$

\begin{figure}[t!]
\centering
\includegraphics[scale=0.75]{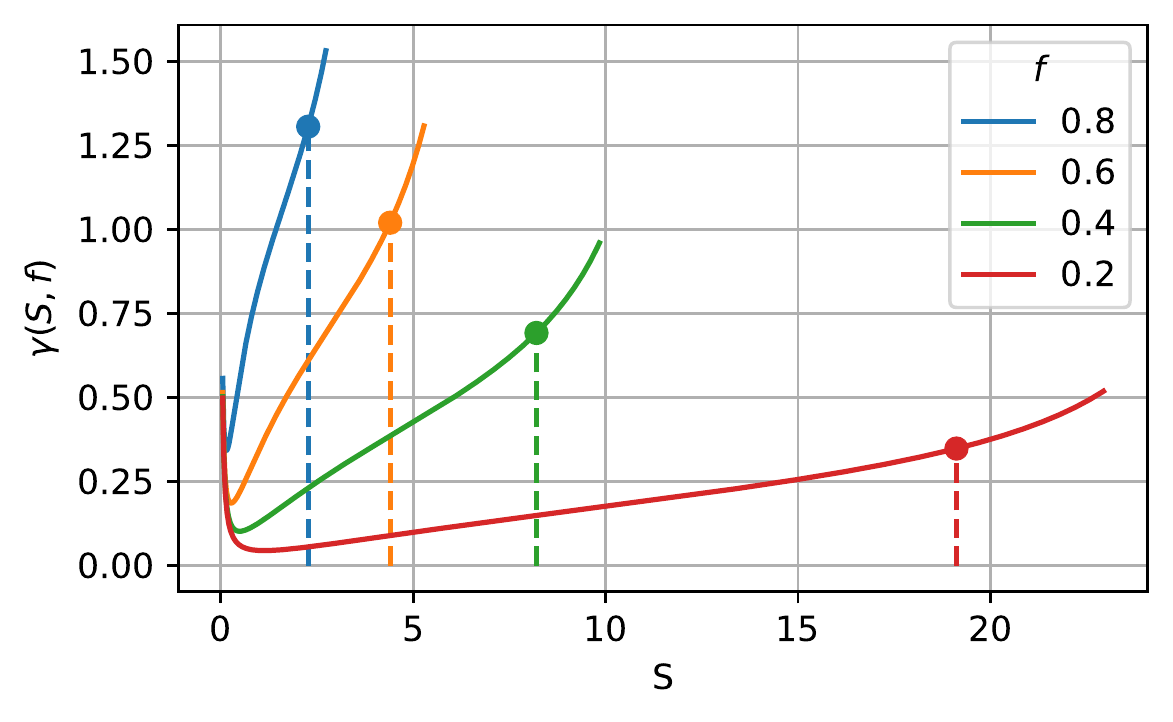}
\caption{Plots of the function $\gamma(S,f)$ found in the slow-dynamics ODE. The dashed vertical lines indicate the values of $S = \Sigma_2(f)$  where the values of $f$ correspond to those in the legend.}\label{fig:gamma_plots}
\end{figure}

Taking the dot product of \eqref{eq:q_i1_system} with $\pmb{\Psi}_c$
and integrating over a disk of radius $R$ gives
\begin{equation}\label{eq:Psi_c-integration}
\int_0^{2\pi} \biggl( \pmb{P}\cdot \frac{\partial \pmb{q}_{i1}}{\partial\rho} - \pmb{q}_{i1}\cdot\frac{\partial \pmb{P}}{\partial\rho}\biggr)\biggr|_{\rho = R} \cos(\omega) R d\omega = -\int_0^R\int_0^{2\pi} \pmb{P}\cdot \pmb{f}_i \cos(\omega)\rho d\rho d\omega.
\end{equation}
As $R\rightarrow\infty$ we calculate
\begin{equation*}
\biggl( \pmb{P}\cdot \frac{\partial \pmb{q}_{i1}}{\partial\rho} - \pmb{q}_{i1}\cdot\frac{\partial \pmb{P}}{\partial\rho}\biggr)\biggr|_{\rho = R} \sim \frac{2}{R}(\cos\omega,\sin\omega)^T \mathcal{J}_i^T \pmb{\alpha}_i,
\end{equation*}
and therefore
\begin{equation*}
\lim_{R\rightarrow\infty} \int_0^{2\pi} \biggl( \pmb{P}\cdot \frac{\partial \pmb{q}_{i1}}{\partial\rho} - \pmb{q}_{i1}\cdot\frac{\partial \pmb{P}}{\partial\rho}\biggr)\biggr|_{\rho = R} \cos(\omega) R d\omega =  2\pi \pmb{e}_1^T \mathcal{J}_i^T \pmb{\alpha}_i,
\end{equation*}
where $\pmb{e}_1 = (1,0)^T$. The right-hand-side of
\eqref{eq:Psi_c-integration} is evaluated by first recalling that
$w_{i0}|_{y_3=0}$ is radially symmetric and hence its contribution
vanishes, whereas
$\nabla_y u_{i0} = u_{i0}'(\rho) (\cos\omega, \sin\omega)^T$ and
therefore
\begin{equation}
\int_0^\infty P_1(\rho) u_{i0}'(\rho)\rho d\rho \int_0^{2\pi} \mathcal{T}_i\cdot (\cos^2\omega,\sin\omega)^T d\omega = \pi\pmb{e}_1\cdot\mathcal{T}_i\int_0^\infty P_1(\rho) u_{i0}'(\rho)\rho d\rho.
\end{equation}
If we instead take the inner product of \eqref{eq:q_i1_system} with
$\pmb{\Psi}_s$ then the computation proceeds identically but with
$\pmb{e}_1$ replaced by $\pmb{e}_2 := (0,1)^T$. Both components of
$\mathcal{T}_i$ are in this way determined and we obtain
$\mathcal{T}_i = \gamma_i\mathcal{J}_i^T\pmb{\alpha}_i$ where
\begin{equation}\label{eq:definition-gamma_i}
\gamma_i := \gamma(S_i;f) = -\frac{2}{\int_0^\infty P_1(\rho)u_{i0}'(\rho)\rho d\rho}.
\end{equation}
The plots of $\gamma(S;f)$ in Figure \ref{fig:gamma_plots} indicate
that $\gamma_i > 0$ (though this awaits a rigorous proof). The
integrals appearing in the definition of $\pmb{\alpha}_i$ can be
calculated using \eqref{eq:integral_grad_Gm_Grm} and
\eqref{eq:integral_grad_Gm_Grb}. Then, using
\begin{equation*}
\mathcal{J}_i^T(\mathbb{I}_3 - x_ix_i^T) = \mathcal{J}_i^T,\quad \mathcal{J}_i^T x_i = 0,\quad \text{and}\quad  \mathcal{J}_i^Tx_j = \begin{pmatrix} -\sin\theta_j\sin(\varphi_i-\varphi_j) \\ \sin\theta_j\cos\theta_i\cos(\varphi_i-\varphi_j) - \sin\theta_i\cos\theta_j\end{pmatrix},
\end{equation*}
we find 
\begin{align*}
\mathcal{J}_i^T\pmb{\alpha}_i = & \frac{1}{2} \sum_{j\neq i}S_j C( x_i^Tx_j) \begin{pmatrix} \sin \theta_j \sin(\varphi_i - \varphi_j) \\ \sin\theta_i\cos\theta_j - \sin\theta_j\cos\theta_i\cos(\varphi_i-\varphi_j) \end{pmatrix} \\
& + \frac{E_0 K_2}{2\pi D_w \sqrt{D_v}}\frac{I_{\nabla G_m}^\perp(\eta_0,x_i^T\hat{x}_0)}{\sqrt{1 - (x_i^T\hat{x}_0)^2}} \begin{pmatrix} \sin \theta_0 \sin(\varphi_i - \varphi_0) \\ \sin\theta_i\cos\theta_0 - \sin\theta_0\cos\theta_i\cos(\varphi_i-\varphi_0) \end{pmatrix},
\end{align*}
where $\hat{x}_0 = x_0 / \eta_0$ has spherical coordinates $(\theta_0,\varphi_0)$ and
\begin{equation}\label{eq:definition-C}
C(z) := \frac{1}{1-z} + \frac{2f}{1-f}\frac{K_1K_2}{D_w}\frac{I_{\nabla G_m}^\perp(1,z)}{\sqrt{1-z^2}}.
\end{equation}
Recalling the definition of $\mathcal{T}_i$ we are confronted with the
system of $2\times N$ ODEs
\begin{subequations}\label{eq:ode-system-all}
\begin{equation}\label{eq:ode-system-spherical}
\begin{split}
\begin{pmatrix} \sin\theta_i\frac{d\varphi_i}{d\sigma} \\ \frac{d\theta_i}{d\sigma}\end{pmatrix} = \gamma_i \biggl\{ & \frac{1}{2} \sum_{j\neq i}S_j C( x_i^Tx_j) \begin{pmatrix} \sin \theta_j \sin(\varphi_i - \varphi_j) \\ \sin\theta_i\cos\theta_j - \sin\theta_j\cos\theta_i\cos(\varphi_i-\varphi_j) \end{pmatrix} \\
& + \frac{E_0 K_2}{2\pi D_w \sqrt{D_v}}\frac{I_{\nabla G_m}^\perp(\eta_0,x_i^T\hat{x}_0)}{\sqrt{1 - (x_i^T\hat{x}_0)^2}} \begin{pmatrix} \sin \theta_0 \sin(\varphi_i - \varphi_0) \\ \sin\theta_i\cos\theta_0 - \sin\theta_0\cos\theta_i\cos(\varphi_i-\varphi_0) \end{pmatrix}\biggr\},
\end{split}
\end{equation}
for each $i=1,...,N$. Equivalently, we can use
$x_i'(\sigma) = \mathcal{J}_i \mathcal{T}_i$ and
\eqref{eq:J_i_identities} to obtain the system of ODEs
\begin{equation}\label{eq:ode-system-euclidean}
\frac{d x_i}{d\sigma} = \gamma_i \bigl(\mathbb{I}_3 - x_ix_i^T\bigr) \biggl\{\frac{1}{2}\sum_{j\neq i} S_j C(x_i^Tx_j)(x_i-x_j) + \frac{E_0K_2}{2\pi D_w\sqrt{D_v}}\frac{I_{\nabla G_m}^\perp(\eta_0,x_i^T\hat{x}_0)}{\sqrt{1-(x_i^T\hat{x}_0)^2}}(x_i - \hat{x}_0)  \biggr\},
\end{equation}
\end{subequations}
for each $i=1,...,N$. The combined system \eqref{eq:NAS} and \eqref{eq:ode-system-all} must be solved simultaneously for the spot strengths $S_1,...,S_N$ and locations $x_1,...,x_N$ and is therefore commonly referred to as a system of Differential-Algebraic-Equations (DAE).

We state the following proposition, whose derivation we deferred to
Appendix \ref{app:proof-prop}, which along with the positivity of
$\gamma_i$ provides a clear geometric interpretation of each term in
\eqref{eq:ode-system-euclidean}.

\begin{proposition}\label{thm:I-C-sign}
  Let $C(\xi)$ be as in \eqref{eq:definition-C} and
  $I_{\nabla G_m}^\perp(\eta_0,\xi)$ as in
  \eqref{eq:I_grad_perp}. Then
\begin{equation}
C(\xi) > 0 \qquad\text{for any}\quad -1<\xi<1\quad\text{and}\quad 0 \leq K_1 < K_1^\star,
\end{equation}
and
\begin{equation}
I_{\nabla G_m}^\perp(\eta_0,\xi) < 0\quad\text{for any }\quad -1<\xi<1\quad\text{and}\quad 0 < \eta_0 < 1.
\end{equation}
\end{proposition}

Since $(\mathbb{I}_3 - x_i x_i^T)$ is a projection onto the tangent
plane of $\partial\Omega$ at $x_i$ we make the following
observations. First, spots are attracted to $\hat{x}_0$, which
coincides with the closest point on $\partial\Omega$ to the
bulk-source term $x_0$. Next, the term appearing in $C(x_i^Tx_j)$ of the
form
\begin{equation*}
\frac{2f}{1-f}\frac{K_1K_2}{D_w}\frac{I_{\nabla G_m}^\perp(1,x_i^Tx_j)}{\sqrt{1-x_i^Tx_j^2}},
\end{equation*}
correspond to an attraction between $x_i$ and $x_j$. However, by Proposition \ref{thm:I-C-sign}, since
$K_1 < K_1^\star$ this attraction is overwhelmed by the repulsion
resulting from the first term in $C(\xi)$. The attractive force
towards the bulk source location is analogous to the attraction
towards the maximum location of an inhomogeneous source explored for
the unit disk in \cite{tzou2018}. However, the attraction between
spots that coupling introduces is novel and is a direct consequence of
the recirculation phenomenon. In the next section we will analyse a
two-spot ring configuration where recirculation plays a key role in
the emergence of steady ``tilted'' configurations. Additionally we
will illustrate a variety of behaviour by integrating the ODE system
\eqref{eq:ode-system-all} for $N=3$ ring configurations.

\section{Examples of the Theory}\label{sec:examples}

In this section we consider a one-, two-, and three-spot ring
configuration. For the one spot configuration we illustrate that the
spot always tends to the location on $\partial \Omega$ closest to
$x_0$. For $N=2$ we will identify the instability thresholds with
respect to both the $O(1)$ eigenvalues as well as the ODE dynamics
\eqref{eq:ode-system-all}. Finally we will highlight some of the
possible behaviours of an $N=3$ spot configuration by
numerically integrating \eqref{eq:ode-system-euclidean}. Without loss
of generality, in this section we will assume
$x_0 = (0,0,\eta_0)^T$ where $0\leq \eta_0 < 1$.

\subsection{One-Spot Configuration}

When $N=1$ the ODE system \eqref{eq:ode-system-euclidean} reduces to
\begin{equation}\label{eq:ode-one-spot}
\frac{d x_1}{d\sigma} = \gamma_1 \bigl(\mathbb{I}_3 - x_1x_1^T\bigr) \frac{E_0K_2}{2\pi D_w\sqrt{D_v}}\frac{I_{\nabla G_m}^\perp(\eta_0,x_1^T\hat{x}_0)}{\sqrt{1-(x_1^T\hat{x}_0)^2}}(x_1 - \hat{x}_0).
\end{equation}

If $\eta_0=0$ then we see that the right-hand-side vanishes
and any point is a stable equilibrium. However, if $\eta_0>0$ then
\eqref{eq:ode-one-spot} has an equilibrium at $(0,0,1)^T$ and at
$(0,0,-1)^T$. Since $I_{\nabla G_m}^\perp(\eta_0,\xi) < 0$ we see that
$(0,0,-1)^T$ is unstable and $(0,0,1)^T$ is globally
attracting. Therefore, if a one-spot solution is stable with respect
to the $O(1)$ competition and splitting instabilities of
\S\ref{sec:stability}, it will concentrate at the point $(0,0,1)$
closest to the source location $x_0$.

\subsection{Two-Spot Configurations}

We first consider the case $\eta_0=0$ individually. Since the matrix
$\mathcal{G}$ is symmetric and $\pmb{g}_{rb}$ is proportional to
$(1,1)^T$, the common source solution $\pmb{S} = S_c\pmb{e}$ solves
the NAS \eqref{eq:NAS} exactly. The ODE system
\eqref{eq:ode-system-euclidean} therefore reduces to
$$
\frac{dx_1}{d\sigma} = \frac{1}{2}\gamma_c S_c\bigl( \mathbb{I} - x_1x_1^T\bigr) C(x_1\cdot x_2)(x_1-x_2),\qquad \frac{dx_2}{d\sigma} = \frac{1}{2}\gamma_c S_c \bigl( \mathbb{I} - x_2x_2^T\bigr) C(x_2\cdot x_1)(x_2-x_1),
$$
where $\gamma_c = \gamma(S_c,f)$. Setting $\cos\beta = x_1^T x_2$ we calculate
$$
\frac{d\beta}{d\sigma} = \frac{1}{2}\gamma_cS_c C(\cos\beta)\sin\beta.
$$
Using the positivity of $C(\xi)$ on $-1<\xi<1$ we deduce that
$\beta = \pi$ is the only stable equilibrium. Therefore the
anti-podal configuration, $x_1 = -x_2$ is the only stable
configuration on a long time-scale when $\eta_0 = 0$.

\begin{figure}[t!]
\centering
\includegraphics[scale=0.75]{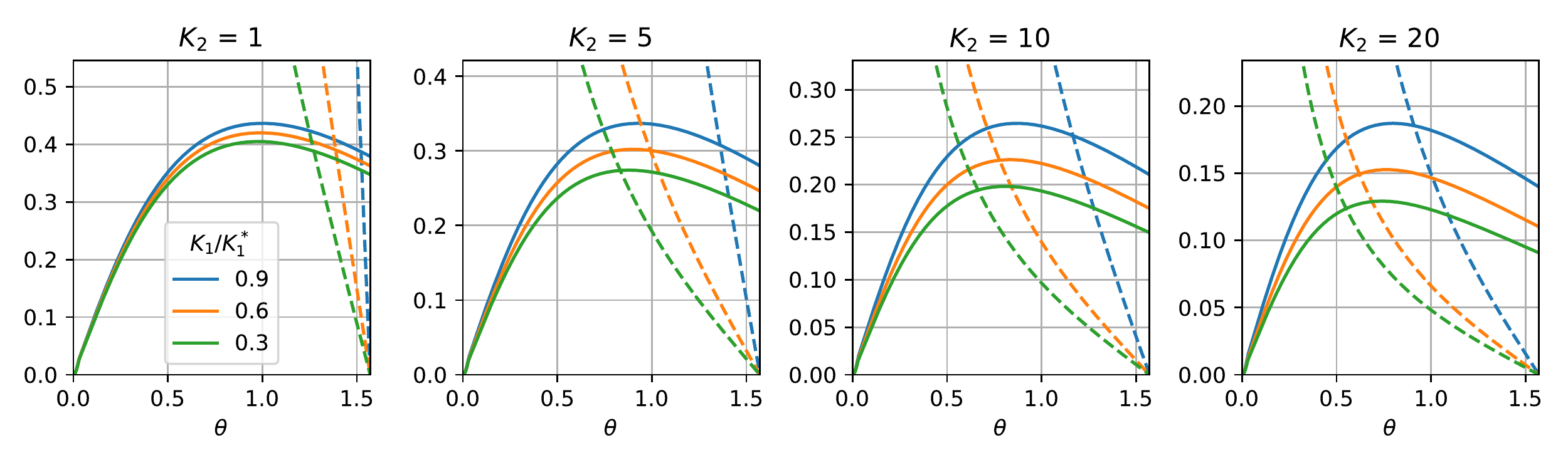}
\caption{Plots of
  $0.25 g_0(\eta_0)\bigl(1 -
  \tfrac{f}{1-f}\tfrac{K_1K_2}{D_w}g_0(1)\bigr)^{-1} C(\cos
  2\theta)\sin 2\theta$ (dashed) and
  $-I_{\nabla G_m}^\perp(\eta_0,\theta)$ (solid) versus $\theta$ for
  various values of $K_2$ and $K_1/K_1^\star$. The common angle
  solution to \eqref{eq:common-ring-equation} corresponds to the
  intersection between the dashed and solid curves. The other
  problem parameters $f=0.4$, $D_w=5$ and $\eta_0=0.6$ are
  fixed.} \label{fig:ring-angle-zero}
\end{figure}

\begin{figure}[t!]
\centering
\includegraphics[scale=0.75]{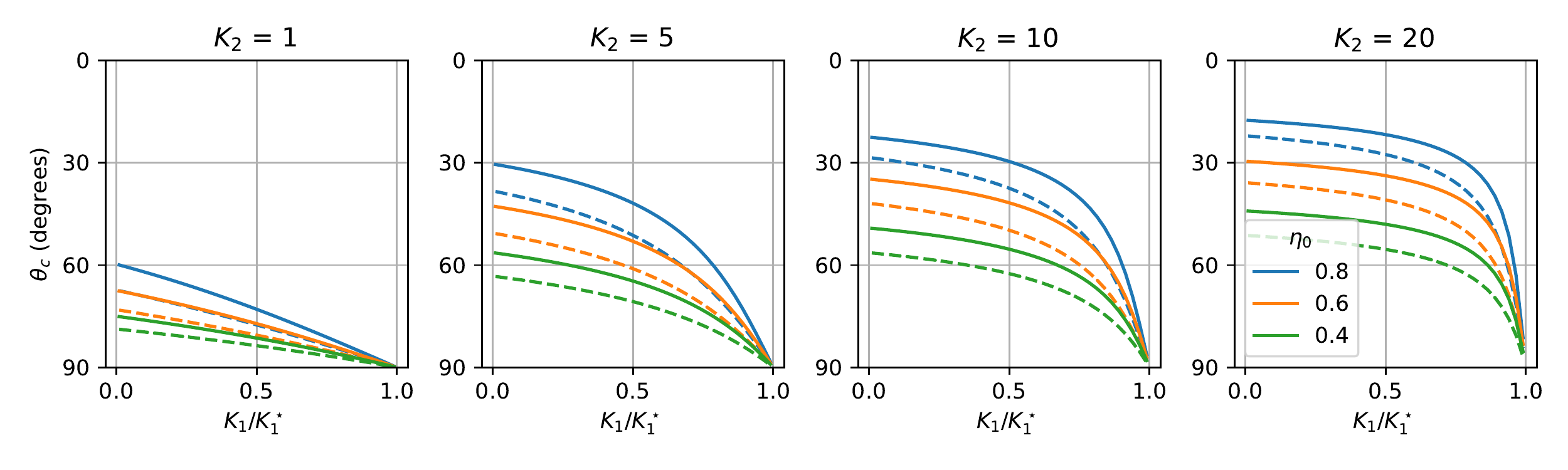}
\caption{Dependence of the common angle for a ring solution consisting
  of $N=2$ (solid) and $N=3$ (dashed) spots as the problem
  parameters $K1$, $K_2$, and $\eta_0$ are varied. The fixed problem
  parameters are $f = 0.4$ and $D_w=5$. }\label{fig:theta_c-2-3-spots}
\end{figure}

We now consider the case $\eta_0 > 0$. From
\eqref{eq:ode-system-euclidean} it is clear that $x_1 = (0,0,1)^T$ and
$x_2 = (0,0,-1)^T$ is an equilibrium configuration. To find other
equilibrium configurations it is instructive to write out the system
\eqref{eq:ode-system-spherical} explicitly
\begin{subequations}
{\small
\begin{equation}\label{eq:ode-two-spots-full}
\begin{split}
\begin{pmatrix} \tfrac{\sin\theta_1}{\gamma_1}\frac{d\varphi_1}{d\sigma} \\ \tfrac{1}{\gamma_1}\frac{d\theta_1}{d\sigma}\end{pmatrix} =  \frac{S_2 C( x_1^Tx_2)}{2} \begin{pmatrix} \sin \theta_2 \sin(\varphi_1 - \varphi_2) \\ \sin\theta_1\cos\theta_2 - \sin\theta_2\cos\theta_1\cos(\varphi_1-\varphi_2) \end{pmatrix}  + \frac{E_0 K_2}{2\pi D_w \sqrt{D_v}}I_{\nabla G_m}^\perp(\eta_0,\cos\theta_1) \begin{pmatrix} 0 \\ 1 \end{pmatrix}, \\
\begin{pmatrix} \tfrac{\sin\theta_2}{\gamma_2}\frac{d\varphi_2}{d\sigma} \\ \tfrac{1}{\gamma_2}\frac{d\theta_2}{d\sigma}\end{pmatrix} =  \frac{S_1 C( x_2^Tx_1)}{2} \begin{pmatrix} \sin \theta_1 \sin(\varphi_2 - \varphi_1) \\ \sin\theta_2\cos\theta_1 - \sin\theta_1\cos\theta_2\cos(\varphi_2-\varphi_1) \end{pmatrix}  + \frac{E_0 K_2}{2\pi D_w \sqrt{D_v}}I_{\nabla G_m}^\perp(\eta_0,\cos\theta_2) \begin{pmatrix} 0 \\ 1 \end{pmatrix}.
\end{split}
\end{equation}}
Assuming neither spot is at $(0,0,\pm 1)$ we will have $-1<x_1^Tx_2<1$ and therefore $\tfrac{d\varphi_1}{d\sigma} = 0$ and $\tfrac{d\varphi_2}{d\sigma} = 0$ only if $\varphi_1 - \varphi_2 = \pm\pi$. Without loss of generality we assume $\varphi_1=0$ and $\varphi_2=\pi$ so that $x_1^Tx_2 = \cos(\theta_1 + \theta_2)$. With this we find that \eqref{eq:ode-two-spots-full} reduces to
\begin{equation}\label{eq:ode-two-spots-reduced}
\begin{split}
\frac{1}{\gamma_1}\frac{d\theta_1}{d\sigma} = \frac{1}{2}S_2C(\cos(\theta_1+\theta_2))\sin(\theta_1+\theta_2)  + \frac{E_0K_2}{2\pi D_w\sqrt{D_v}}I_{\nabla G_m}^\perp(\eta_0,\cos\theta_1),\\
\frac{1}{\gamma_2}\frac{d\theta_2}{d\sigma} = \frac{1}{2}S_1C(\cos(\theta_1+\theta_2))\sin(\theta_1+\theta_2)  + \frac{E_0K_2}{2\pi D_w\sqrt{D_v}}I_{\nabla G_m}^\perp(\eta_0,\cos\theta_2).
\end{split}
\end{equation}
\end{subequations}
We seek a common angle equilibrium solution to
\eqref{eq:ode-two-spots-reduced} by setting
$\theta_1=\theta_2=\theta_c$. Since $x_1$ and $x_2$ make a common
angle $\theta_c$ with the bulk source location $x_0$ we find that
$\pmb{S} = S_c\pmb{e}$ solves the NAS \eqref{eq:NAS} exactly. The
common angle is then found by solving
\begin{equation}\label{eq:common-ring-equation}
\frac{1}{4}\frac{g_0(\eta_0)}{1 - \tfrac{f}{1-f}\tfrac{K_1K_2}{D_w}g_0(1)} C(\cos 2\theta_c)\sin 2\theta_c + I_{\nabla G_m}^\perp(\eta_0,\cos\theta_c) = 0,
\end{equation}
which we remark is independent of the bulk-source strength $E_0$ and
membrane diffusivity $D_v$. By the positivity of $C(\xi)$ and the
negativity of $I_{\nabla G_m}^\perp(\eta_0,\xi)$ (see Proposition \ref{thm:I-C-sign}) we deduce that the
common angle must be in the interval $0 < \theta_c < \pi /
2$. Moreover, since $C(\cos 2\theta)\sin 2\theta$ diverges to
$+\infty$ as $\theta\rightarrow 0^+$ the left-hand-side of
\eqref{eq:common-ring-equation} changes sign on the interval
$0<\theta_c<\pi/2$ and so a solution must exist. As indicated by
Figure \ref{fig:ring-angle-zero}, our numerical calculations further
indicate that this solution is unique. We find $\theta_c$ by solving
\eqref{eq:common-ring-equation} numerically.
In Figure \ref{fig:theta_c-2-3-spots} the resulting
dependence of $\theta_c$ on the model parameters is illustrated. 

Next we investigate the stability of the common angle ring solution
constructed above. The splitting instability threshold is given
explicitly by \eqref{eq:splitting-threshold-common} while the
competition threshold is determined by \eqref{eq:competition-A}. We
note that $k^\star$ is the eigenvalue corresponding to the eigenvector
$(1,-1)^T$ of the $2\times 2$ matrix $\mathcal{G}$ and is thus given
by $\mathcal{G}_{11}-\mathcal{G}_{12}$. The threshold $K_{1C}$ is then
determined by numerically solving \eqref{eq:competition-A}, where we
use \eqref{eq:competition-leading-order} to aid the root- finding
algorirthm.

To determine the stability threshold with respect to the slow dynamics
we linearize \eqref{eq:ode-two-spots-full} about the common angle
solution $(\varphi_1,\theta_1) = (0,\theta_c)$ and
$(\varphi_2,\theta_2)=(\pi,\theta_c)$. The rotational symmetry about
the $z$-axis implies that the ring configuration is neutrally stable
to perturbations of the form
$(\varphi_1,\varphi_2) = (\delta,\pi+\delta)$ while it is
unconditionally stable with respect to any other perturbations in
$(\varphi_1,\varphi_2)$ because of the mutual repulsion between
spots. The stability of the ring solution is therefore determined
solely by its response to perturbations in $\theta_1$ and
$\theta_2$. We define $F_1(\theta_1,\theta_2)$ and
$F_2(\theta_1,\theta_2)$ by the right-hand-sides of
\eqref{eq:ode-two-spots-reduced}. We must then calculate the
eigenvalues of the $2\times 2$ matrix
$\bigl( \partial_{\theta_j} F_i
\bigr)\bigr|_{\theta_1=\theta_2=\theta_c}$. This requires that we
first calculate the derivatives of $\pmb{S}$ with respect to
$\theta_1$ and $\theta_2$. Using the eigenvectors
$\pmb{q}_1 = (1,1)^T$ and $\pmb{q}_2 = (1,-1)^T$ of $\mathcal{G}$ we
write
\begin{equation*}
\pmb{S} = \frac{S_1+S_2}{2}\pmb{q}_1 +  \frac{S_1-S_2}{2}\pmb{q}_2,
\end{equation*}
with which the NAS \eqref{eq:NAS} becomes
$$
\pmb{S} + 2\pi\nu \tfrac{S_1-S_2}{2}k^\star \pmb{q}_2 + \nu\bigl(\mathbb{I}_2 - \mathcal{E}_2\bigr)\pmb{\chi}(\pmb{S}) = S_c \pmb{e} + \nu\frac{ E_0 K_2}{D_w\sqrt{D_v}}\bigl(\mathbb{I}_2-\mathcal{E}_2\bigr)\pmb{g}_{rb}.
$$
Differentiating this with respect to $\theta_i$ and evaluating at
$\theta_1=\theta_2=\theta_c$ we find
\begin{equation}\label{eq:two-spot-NAS}
\frac{\partial\pmb{S}}{\partial\theta_i}\biggr|_{\theta_c} + \pi\nu\biggl(\frac{\partial S_1}{\partial\theta_i} - \frac{\partial S_2}{\partial\theta_i}\biggr) k^\star\pmb{q}_2 + \frac{1}{2}\nu\chi'(S_c)\biggl(\frac{\partial S_1}{\partial\theta_i} - \frac{\partial S_2}{\partial\theta_i}\biggr)\pmb{q}_2 = \frac{\nu E_0 K_2}{2 D_w\sqrt{D_v}}\biggl( \frac{\partial g_{rb}(\theta_1)}{\partial\theta_i}\biggr|_{\theta_c} - \frac{\partial g_{rb}(\theta_2)}{\partial\theta_i}\biggr|_{\theta_c}\biggr)\pmb{q}_2,
\end{equation}
where
$$
g_{rb}(\theta) = \int_{\partial\Omega} G_m((\sin\theta,0,\cos\theta)^T,\xi)G_{rb}(\xi,x_0)dA_\xi .
$$
Left multiplying \eqref{eq:two-spot-NAS} by $\pmb{q}_1 = \pmb{e}^T$ we
find $\partial_{\theta_i} S_1 + \partial_{\theta_i} S_2 = 0$. On the
other hand, if we left-multiply by $\pmb{q}_2^T$ then we determine
\begin{equation}
\frac{\partial S_1}{\partial \theta_1} = \frac{\partial S_2}{\partial \theta_2} = -\frac{\partial S_1}{\partial\theta_2} = -\frac{\partial S_2}{\partial \theta_1} = \frac{E_0 K_2}{2 D_w\sqrt{D_v}}\frac{g_{rb}'(\theta_c)}{\mathcal{A}_c^\star},
\end{equation}
where we point out that $\mathcal{A}_c^*$ vanishes at the competition
instability threshold. Next, since the matrix
$\bigl( \partial_{\theta_j} F_i
\bigr)\bigr|_{\theta_1=\theta_2=\theta_c}$ is symmetric and of
constant row sum we immediately find its two eigenvectors $(1,1)^T$
and $(1,-1)^T$ with corresponding eigenvalues given by
\begin{align*}
& \mu_+ = -S_c \sin2\theta_c \frac{d}{dz}\biggr|_{z = \cos 2\theta_c}\sqrt{1-z^2}C(z) - \frac{E_0 K_2}{2\pi D_w \sqrt{D_v}} \frac{\partial I_{\nabla G_m}^\perp}{\partial z}(\eta_0,z)\biggr|_{z=\cos\theta_c}\sin\theta_c,\\
& \mu_- =  -\frac{E_0 K_2}{2 D_w\sqrt{D_v}}\frac{g_{rb}'(\theta_c)}{\mathcal{A}_c^\star} C(\cos2\theta_c)\sin2\theta_c - \frac{E_0 K_2 }{2\pi D_w \sqrt{D_v}} \frac{\partial I_{\nabla G_m}^\perp}{\partial z}(\eta_0,z)\biggr|_{z=\cos\theta_c}\sin\theta_c.
\end{align*}
Since our numerics suggest that $\theta_c$ is the unique solution to
\eqref{eq:common-ring-equation} we assume that $\mu_+ < 0$ for all
parameter values with $K_1 < K_1^\star$ and focus only on
determining the sign $\mu_-$. First we use \eqref{eq:I_Gm_derivative}
and \eqref{eq:I_grad_Gm-ode} to simplify
\begin{equation}
\mu_- = -\frac{E_0 K_2}{D_w\sqrt{D_v}}\biggl[ \frac{1}{4\pi}\biggl(\frac{C(\cos2\theta_c)\sin2\theta_c}{\mathcal{A}_c^\star} + 2\cot\theta_c\biggr)I_{\nabla G_m}^\perp(\eta_0,\cos\theta_c) + G_{rb}(\cos\theta_c,\eta_0) - \frac{1}{4\pi}g_0(\eta_0) \biggr],
\end{equation}
and then numerically solve $\mu_- = 0$ for $K_1$ as a function of
$K_2$ to obtain the threshold $K_{1T}(K_2)$. Our numerical
computations indicate that $\mu_- < 0 $ (resp. $\mu_- > 0$) when
$K_1 < K_{1T}$ (resp. $K_1 > K_{1T}$) and therefore as $K_1$ is
increased beyond this threshold the two-spot ring solution becomes
unstable with respect to the slow dynamics. Since $\mu_-$ corresponds
to the eigenvector $(1,-1)^T$ of the linearization of
\eqref{eq:ode-two-spots-reduced} we expect this instability to result
in a new spot configuration with $\theta_1 > \theta_c$ and
$\theta_2 < \theta_c$.

\begin{figure}[t!]
\centering
\includegraphics[scale=0.75]{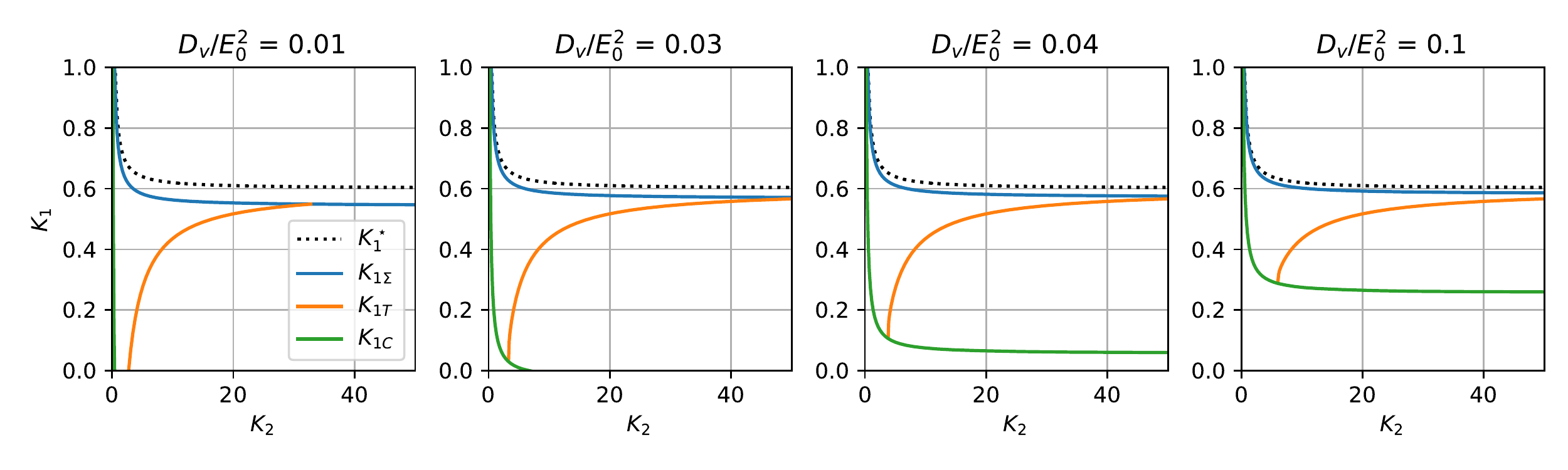}
\caption{Sample bifurcation diagrams for a two-spot common angle ring
  solution. Regions below the green lines (resp. above the blue lines)
  correspond to parameter values where the two spot common angle ring
  solution goes unstable in $O(1)$ time with respect to the
  competition (resp. splitting) instabilities. In the region above the
  orange line the ring solution becomes unstable with respect to the
  ``tilt'' instability in $O(\varepsilon^{-2})$ time. Fixed model
  parameters are $f=0.4$, $D_w=5$, $\eta_0 = 0.6$ and
  $\nu = 0.01$.}\label{fig:two-spot-bifurcation-diagrams}
\end{figure}

\begin{figure}[t!]
\centering
\begin{subfigure}{0.5\textwidth}
	\centering
	\includegraphics[scale=0.75]{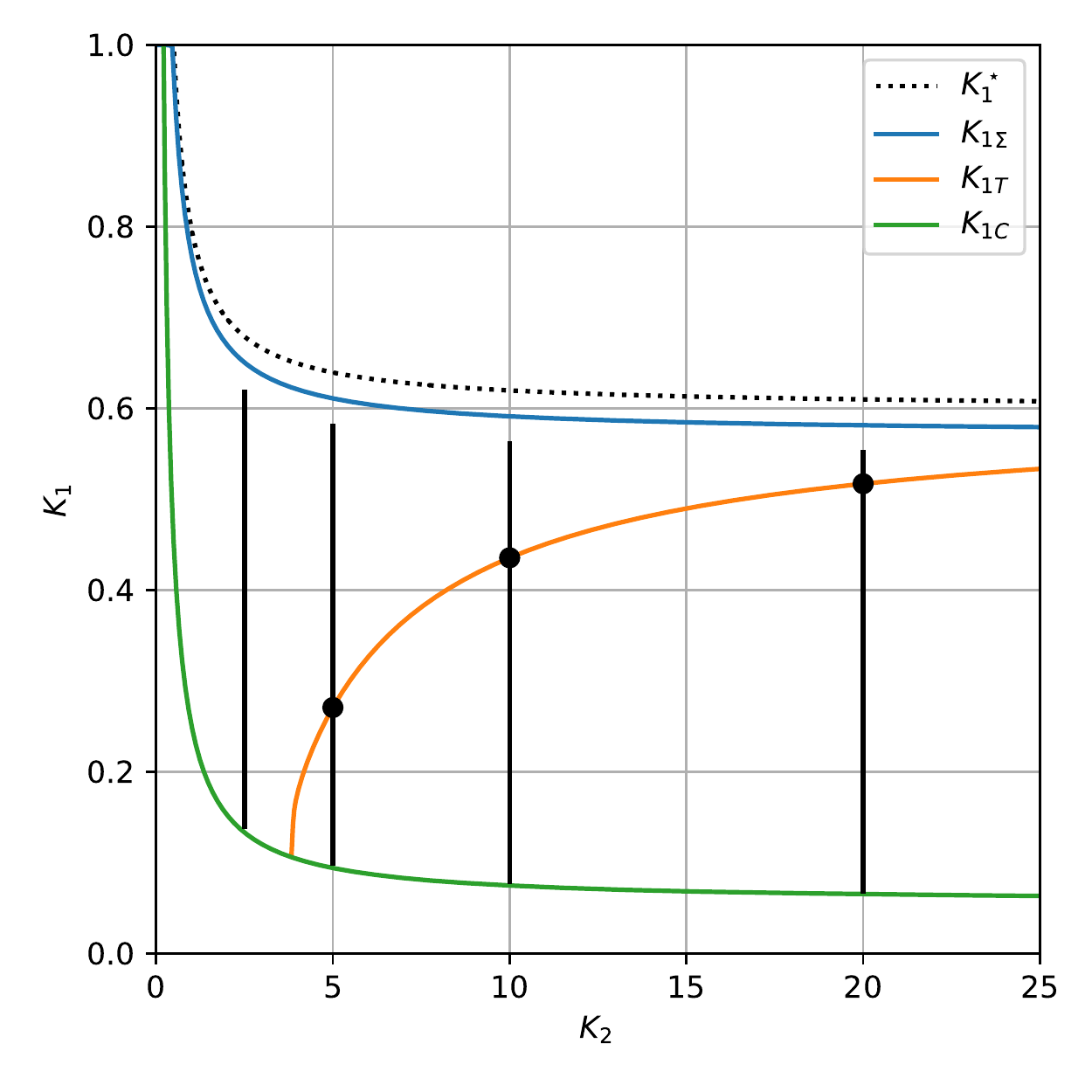}
	\caption{}
\end{subfigure}%
\begin{subfigure}{0.5\textwidth}
	\centering
	\includegraphics[scale=0.75]{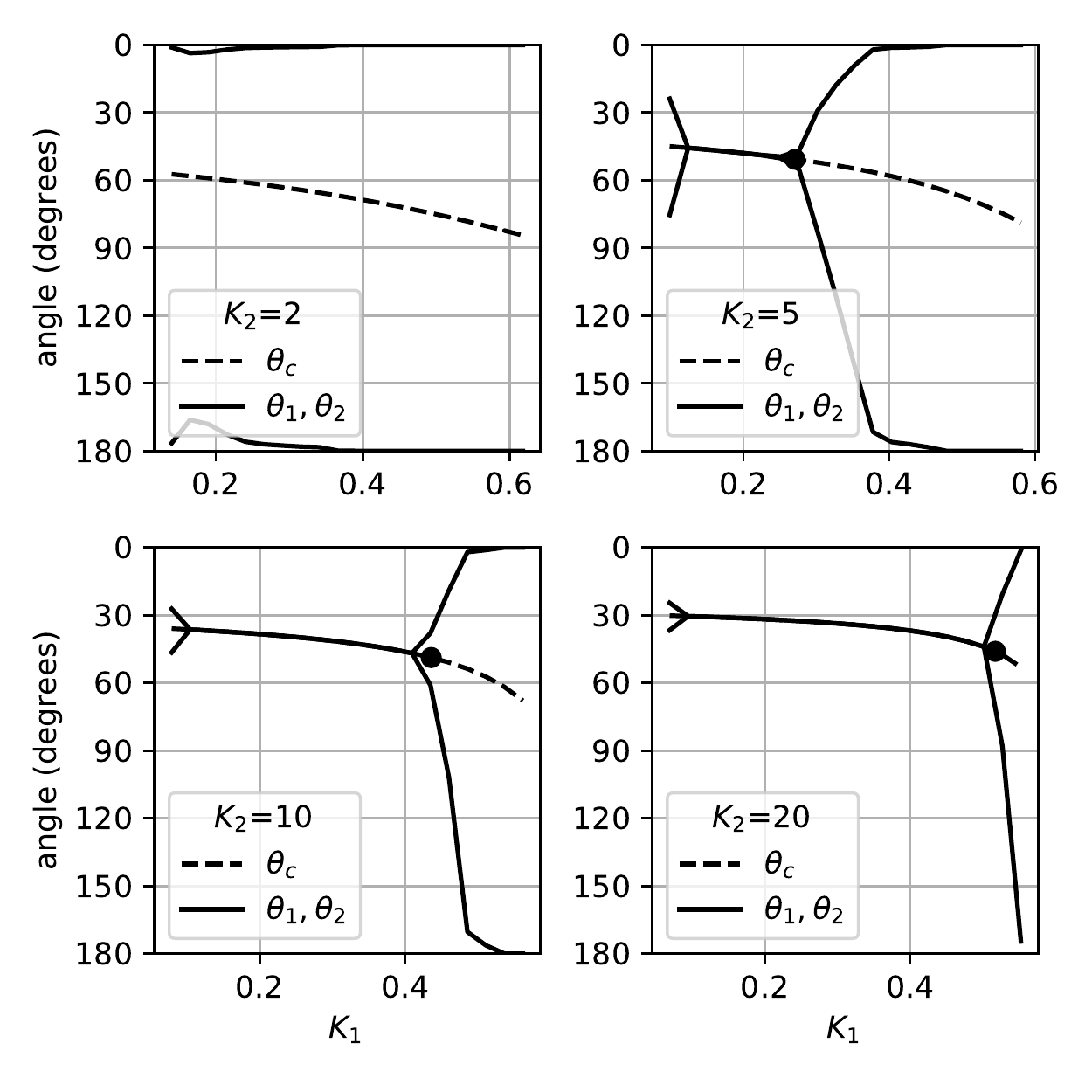}
	\caption{}
\end{subfigure}%
\caption{(a) Bifurcation diagram for a two-spot common angle solution,
  and (b) equilibrium solutions of the slow-dynamics ODE
  \eqref{eq:ode-two-spots-full} as $K_1$ is varied beyond the $K_{1T}$
  threshold at fixed values of $K_2$ indicated by vertical black lines
  in (a). Fixed problem parameters are $D_v/E_0^2 = 0.04$, $f=0.4$,
  $D_w=5$, $\eta_0 = 0.6$ and $\nu = 0.01$. Equilibrium solutions in
  (b) were obtained by performing a long-time numerical integration of
  \eqref{eq:ode-two-spots-full} with a small perturbation from the
  common angle solution as the initial condition. The tilted branch
  appearing for small values of $K_1$ in (b) are ignored due to the
  proximity of $K_1$ to the competition instability threshold
  $K_{1C}$.}\label{fig:two-spot-bifurcation-diagram-with-continuation}
\end{figure}

In Figure \ref{fig:two-spot-bifurcation-diagrams} we illustrate the
preceding discussion by superimposing the splitting, competition, and
tilt instability thresholds. We remark that in the region bounded by
the competition (green), splitting (blue), and tilt (orange)
instability thresholds, the two-spot ring solution is stable with
respect to all $O(1)$ instabilities, but will undergo a ``tilt''
instability in $O(\varepsilon^{-2})$ time. Our previous discussion
suggests that within this region a new tilted, or asymmetric, stable
two-spot configuration should exist. By numerically continuing the
common angle solution from $K_1 < K_{1T}$ into the region where
$K_1 > K_{1T}$ our results in Figure
\ref{fig:two-spot-bifurcation-diagram-with-continuation} illustrate
the emergence of these new types of solutions, and furthermore suggest
that as $K_1$ continues to increase towards $K_{1\Sigma}$ one angle
will tend to $0$ and the other to $\pi$. Finally, note that we expect
the tilted solutions to be stable with respect to the $O(1)$
instabilities provided $K_1$ is sufficiently far from $K_{1C}$ and
$K_{1\Sigma}$ since $\pmb{S}$ will be an $O(\nu)$ perturbation away
from the common source solution $S_c\pmb{e}$ to which the plotted
thresholds correspond.

\subsubsection{Three-Spot Configuration}

We conclude this section with some illustrative examples of the
dynamics obtained by solving \eqref{eq:ode-system-all} for $N=3$. In
such a case there is more diversity in the possible equilibrium
solutions and the ways in which they can become unstable. We will
forego the more detailed analysis we performed for $N=2$ and instead
numerically integrate \eqref{eq:ode-system-all} for small
perturbations away from a common angle solution. It is worth noting
that there are many questions left unanswered for the dynamics of
\eqref{eq:ode-system-all} when $N\geq 2$ even for the uncoupled case
(see \S 5 of \cite{trinh2016} for a more detailed display of dynamics
and open problems).

\begin{figure}[t!]
	\centering
	\includegraphics[scale=0.75]{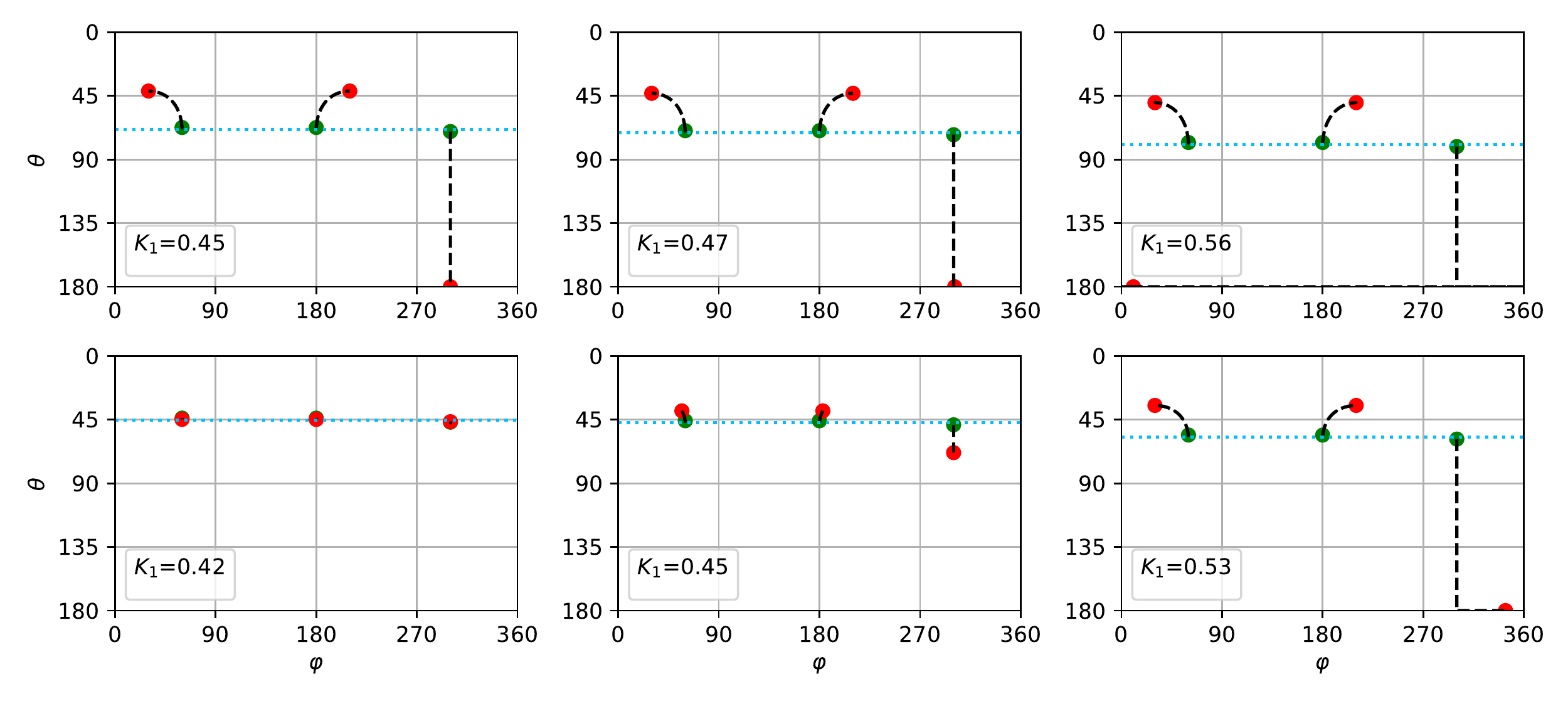}
	\caption{Numerically computed spot dynamics starting from an up-up-down perturbation away from the three-spot common angle ring solution for $K_2 = 5$ (top row) and $K_2=20$ (bottom row). Green dots indicate the starting configuration and red dots correspond to the final spot configuration while the dashed horizontal line indicates the common angle of the three-spot ring solution. The remaining problem parameters are fixed and given by $f=0.4$, $D_w=5$, $\eta_0=0.6$, $D_v/E_0^2 = 0.04$, and $\nu = 0.01$.}\label{fig:uud_plot}
\end{figure}

\begin{figure}[t!]
\centering
\includegraphics[scale=0.75]{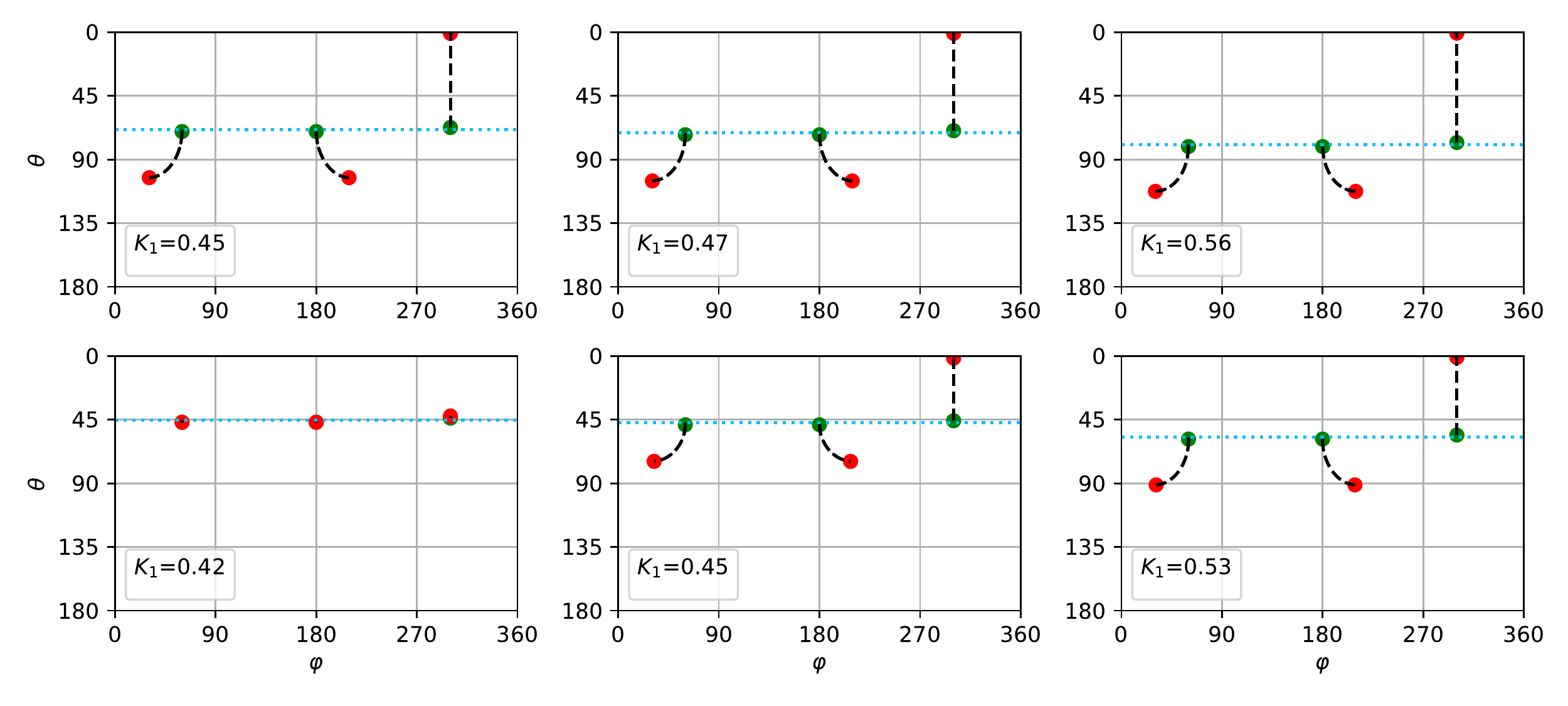}
\caption{Numerically computed spot dynamics starting from a down-down-up perturbation away from the three-spot common angle ring solution for $K_2 = 5$ (top row) and $K_2=20$ (bottom row). Green dots indicate the starting configuration and red dots correspond to the final spot configuration while the dashed horizontal line indicates the common angle of the three-spot ring solution. The remaining problem parameters are fixed and given by $f=0.4$, $D_w=5$, $\eta_0=0.6$, $D_v/E_0^2 = 0.04$, and $\nu = 0.01$.}\label{fig:ddu_plot}
\end{figure}

We begin by constructing the common-angle three-spot ring solution by setting
\begin{equation}\label{eq:three-spot-configuration}
x_i = (\sin\theta_c\cos\varphi_i,\sin\theta_c\sin\varphi_i,\cos\theta_c)^T,\qquad\text{where}\quad \varphi_i = \tfrac{2\pi (i-1)}{N}\qquad (i=1,2,3),
\end{equation}
where $\theta_c$ is to be determined. From symmetry considerations one
can show that $d\varphi_1/dt = d\varphi_2/dt = d\varphi_3/dt = 0$ for
such a configuration and setting either of the remaining equations to
zero yields an equation for $\theta_c$. The parameter dependence of
the common angle for this three-spot ring solution can be found in
Figure \ref{fig:theta_c-2-3-spots}. Notice that the three-spot angle
is greater than the corresponding two-spot angle. In Figures
\ref{fig:uud_plot} and \ref{fig:ddu_plot} we
illustrate some of the possible dynamics by integrating
\eqref{eq:ode-system-all} starting with two types of perturbations
away from the common angle ring solution. In the first we perturb two
spots upward and one spot downward (UUD), while in the second we perturb two spots downward and own spot upward (DDU). In these Figures we observe that
for $K_2 = 5$ the ring solution in both cases immediately becomes
unstable. For the UUD (resp. DDU) perturbation, the resulting
configuration consists of a two-spot ring at an angle greater than
(resp. less than) the three-spot ring angle, and the remaining spot
tending towards the south (resp. north) pole. On the
other hand at $K_2 = 20$ we observe that the ring solution only goes
unstable after $K_1$ exceeds some threshold. The final configurations
after the instability has been triggered remain qualitatively the same
as those for $K_2=5$. Moreover comparing Figures \ref{fig:uud_plot} and \ref{fig:ddu_plot} we observe out that a smaller value of
$K_1$ appears to be needed to make the ring solution unstable when we apply a DDU instead of UUD perturbation.

\section{Discussion}\label{sec:discussion}

In this paper we considered a bulk-membrane coupled model that
consists of a reaction diffusion system with Brusselator kinetics on
the unit sphere coupled to a bulk diffusion process within the unit
ball. Specifically we have assumed that the inhibitor is bound to the
membrane, whereas the activator occupies both the cytosol, where it is
generated and undergoes bulk diffusion, and the membrane, where it
diffuses and reacts with the inhibitor. Additionally, we have chosen
to describe the attachment-detachment process for the activator using
a linear Robin boundary condition. One of the primary motivations for
considering this model is to give the feed rate found in the uncoupled
Brusselator model a clear origin. In our model the feed rate results
from an activator point source within the bulk, modelled by a Dirac
delta, which diffuses outward towards the membrane. This on its own
does not lead to new coupling-dependent behaviour since we could have
equally well considered an uncoupled model with a heterogeneous feed
term. However, our choice of coupling also allows the membrane bound
activator to detach from the membrane and enter the bulk, after which
it may reattach to the membrane. This \textit{recirculation} effect
leads to new results regarding the existence, stability, and slow
dynamics of localized spot patterns.

Our analysis focused on the singularly perturbed limit where the
activator diffusivity is given by an asymptotically small
parameter. Using matched asymptotic expansions we constructed a
quasi-equilibrium solution consisting of $N$ strongly localized spots
arranged on the unit sphere. Our asymptotic analysis demonstrates that
such $N$-spot patterns can only be constructed if the coupling parameter
$K_1$ does not exceed the threshold $K_1^\star$. If this constraint is
satisfied then we can asymptotically construct the quasi-equilibrium
pattern by solving a system of nonlinear algebraic systems for the
spot strengths $S_1,...,S_N$. Next, we considered the linear stability
of an $N$-spot pattern on an $O(1)$ time scale. Linearizing about the
$N$-spot quasi-equilibrium solution leads to an eigenvalue problem
admitting localized eigenfunctions. Using the method of matched
asymptotics we derived criteria for the stability with respect to
``competition'' and ``splitting instabilities'' previously studied in
detail for the uncoupled Brusselator model in
\cite{rozada2014}. Finally, the stability on an $O(\varepsilon^{-2})$
timescale was analysed by deriving an ODE describing the slow dynamics
of the spot locations. By analysing this ODE in detail for a two-spot
ring configuration we found that spots can undergo a ``tilt'' instability
leading to new asymmetric two-spot configurations that are stable on
both an $O(1)$ and an $O(\varepsilon^{-2})$ time-scale. Further
numerical simulations indicate that similar phenomenon can be obtained
for three-spot patterns.

We conclude with suggestions for future work. First, our
asymptotically derived instability thresholds require numerical
verification by solving the entire PDE system \eqref{eq:pde-all}. We remark here that the quasi-equilibrium construction and stability analysis of \S \ref{sec:quasi-equilibrium} and \S \ref{sec:stability} is accurate to all orders in $\nu$ and their numerical verification therefore requires only small values of $\varepsilon$ with no restrictions on $\nu=-1/\log\varepsilon$. However the long-time integration of the full PDE system \eqref{eq:pde-all} required to verify the DAE \eqref{eq:ode-system-all} provides a significant numerical challenge. This is true also for the DAE derived in \cite{trinh2016} for the uncoupled Brusselator model (see also open problems in \cite{jamieson0216}).

There are several extensions to our analysis and to the model
considered which can be undertaken. First, we have neglected the
$O(1)$ instabilities that may arise through a H\"opf bifurcation. A
previous study of a two-dimensional bulk-membrane coupled model with
Gierer-Meinhardt kinetics revealed a rich dependence of the H\"opf
instability threshold on the coupling parameters as well as the bulk-
and membrane- time constants \cite{gomez2019}. Determining the
coupling dependence of this threshold for our current model may be a
fruitful direction for future work. In addition, extending our model
to include a bulk-bound inhibitor, satisfying possibly nonlinear
kinetics, would lead to a system more similar to that found in the
numerical studies of Madzvamuse et. al.\@
\cite{madzvamuse_2015,madzvamuse_2016} for which a detailed nonlinear analysis
remains to be done. Finally, the role of geometry on the stability and
slow-spot dynamics for coupled (and uncoupled) models remains largely
untouched. The biggest hurdles in this direction include the numerical
evaluation of Green's functions for the Laplace-Beltrami operator on
an arbitrary manifold.

\section*{Acknowledgements}

I would like to thank Prof. Michael Ward and Prof. Juncheng Wei for their thoughtful guidance, support, and many  helpful comments throughout this project. This work was supported by an NSERC CGS-D Doctoral award.

\section{A Scaling of the Coupled Brusselator Model}\label{app:scaling}

We perform a formal scaling of the system \eqref{eq:pde-unscaled} such
that it exhibits strongly localized solutions. Notice that we have
already assumed that $\Omega$ is the unit ball in $\mathbb{R}^3$ which
can be done without loss of generality since any spatial scaling can
be absorbed into the parameters. We will further impose that only the
feed bulk source strength $\mathcal{E}_0$ may depend on the small
parameter $\varepsilon_0$.

The underlying assumption for strongly localized patterns is that $U$
exhibits two distinct scalings $U_i$ and $U_o$ in the regions near and
far away from a spot respectively. On the other hand the inhibitor
exhibits a single global scaling $V_g$. We assume in addition that the
bulk-bound activator exhibits an outer scaling $U_{Bo}$. First, each
spot is localized in an $O(\varepsilon_0)$ region where the Laplacian
will scale like $1/\varepsilon_0^2$. Requiring that $U$ and $V$
interact within the inner scale leads to $O(U_i) = O(U_i^2 V_g)$ by
balancing \eqref{eq:pde-unscaled-U} and therefore $V_g =
O(U_i^{-1})$. Balancing the inner limit of \eqref{eq:pde-unscaled-V}
determines that $O(\varepsilon_0^{-2} V_g) = O(U_i)$ and thus
$$
U_i = O(\varepsilon_0^{-1}),\qquad V_g = O(\varepsilon_0).
$$
Then, balancing equation \eqref{eq:pde-unscaled-V} in the outer region
implies $U_o = O(V_g) = O(\varepsilon_0)$. Turning now to the bulk
equation we observe that to balance the Dirac delta term with $U_B$ in
\eqref{eq:pde-unscaled-UB} we need $O(U_{Bo}) = O(\mathcal{E}_0)$. By
then balancing \eqref{eq:pde-unscaled-U} in the outer region we deduce
that $\mathcal{E}_0 = O(U_o) = O(\varepsilon_0)$. We therefore rescale
the time $T$ and concentrations $U$, $V$, and $U_B$ according to
\begin{equation}
T = \sigma T,\qquad U = \frac{\mu}{\varepsilon_0}u,\qquad V = \nu \varepsilon_0 v,\qquad U_B = \omega \varepsilon_0 w,\qquad  \mathcal{E}_0 = \omega \varepsilon_0 k_B E_0.
\end{equation}
Setting
\begin{equation}
\sigma := \frac{1}{1 + B + \gamma_{\partial\Omega}K_1},\qquad \mu = B,\qquad \nu = 1,\qquad \omega = \frac{B}{k_B}\frac{\gamma_{\Omega}}{\gamma_{\partial\Omega}},
\end{equation}
and defining new coupling and diffusivity parameters
\begin{equation}
K_1 := \frac{\gamma_{\partial\Omega}\mathcal{K}_1}{1+B+\gamma_{\partial\Omega}\mathcal{K}_1},\quad K_2 := \frac{\gamma_\Omega\mathcal{K}_2}{k_B},\quad D_w := \frac{D_B}{k_B},\quad D_v := \frac{D_v(1+B+\gamma_{\partial\Omega}\mathcal{K}_1)}{B^2},
\end{equation}
time constants
\begin{equation}
\tau_v := \frac{(1 + B + \gamma_{\partial\Omega})^2}{B^2},\quad \tau_w := \frac{1 + B + \gamma_{\partial\Omega}K_1}{k_B}
\end{equation}
and
\begin{equation}
f := \frac{B}{1+B+\gamma_{\partial\Omega}K_1},
\end{equation}
we obtain the system \eqref{eq:pde-all}.

\section{Derivation of Proposition \ref{thm:I-C-sign}}\label{app:proof-prop}

To determine the sign of $I_{\nabla G_m}^\perp(\eta_0,z)$ for $-1<z<1$
we first determine an ODE that it satisfies. Differentiating and
using properties of Legendre polynomials we compute
\begin{align*}
\frac{d I_{\nabla G_m}^\perp(\eta_0,z)}{d z} & = - \frac{d}{dz}\biggl( \frac{1}{2\sqrt{1-z^2}}\sum_{l=1}^\infty \frac{g_l(\eta_0)}{l(l+1)}(1-z^2)\frac{dP_l}{dz}   \biggr) \\
& = -\frac{z}{2(1-z^2)^{3/2}}\sum_{l=1}^\infty \frac{g_l(\eta_0)}{l(l+1)}(1-z^2)\frac{d P_l}{dz} - \frac{1}{2\sqrt{1-z^2}}\sum_{l=1}^\infty \frac{g_l(\eta_0)}{l(l+1)}\frac{d}{dz}\biggl[(1-z^2)\frac{dP_l}{dz}\biggr] \\
& = \frac{1}{2\sqrt{1-z^2}}\biggl[ \frac{2z}{\sqrt{1-z^2}}I_{\nabla G_m}^\perp(\eta_0,z) + 4\pi G_{rb}(z,\eta_0) - g_0(\eta_0) \biggr].
\end{align*}
The resulting ODE
\begin{equation*}\label{eq:I_grad_Gm-ode}
\frac{d I_{\nabla G_m}^\perp(\eta_0,z)}{d z} - \frac{z}{1-z^2}I_{\nabla G_m}^\perp(\eta_0,z) = \frac{2\pi}{\sqrt{1-z^2}}\biggl\{ G_{rb}(z,\eta_0) - \frac{1}{4\pi}g_0(\eta_0)\biggr\},
\end{equation*}
can be solved explicitly using $I_{\nabla G_m}^\perp(\eta_0,-1)=0$ to get
\begin{equation}\label{eq:I_grad_Gm-spherical-cap}
I_{\nabla G_m}^\perp(\eta_0,z) = \frac{2\pi}{\sqrt{1-z^2}}\int_{-1}^z \biggl( G_{rb}(z,\eta_0) - \frac{1}{4\pi}g_0(\eta_0) \biggr)dz.
\end{equation}
where we remark that
$$
\frac{1}{4\pi} g_0(\eta_0) = \frac{1}{2}\int_{-1}^1 G_{rb}(z,\eta_0)dz.
$$
Since $G_{rb}(z,\eta_0)$ is monotone increasing in $z$ for $\eta_0>0$
we readily see from \eqref{eq:I_grad_Gm-spherical-cap} that
$I_{\nabla G_m}^\perp(\eta_0,z) < 0$ on $-1<z<1$.

To determine the sign of $C(z)$ we follow a similar procedure and first calculate
$$
\frac{d}{dz}\bigl[(1-z^2)C(z)\bigr] = 1 + \frac{f}{1-f}\frac{K_1K_2}{D_w}\sum_{l=1}^\infty g_l(1) P_l(z),
$$
where have used the differential equation satisfied by $P_l(z)$. Rearranging the sum and recalling \eqref{eq:G_rm_series} we obtain
\begin{equation}
\frac{d}{dz}\bigl[(1-z^2)C(z)\bigr] = 1 - \frac{f}{1-f}\frac{K_1K_2}{D_w}g_0(1) + 4\pi \frac{f}{1-f}\frac{K_1K_2}{D_w}G_{rm}(x,e_z) > 0,\qquad\text{for }K_1<K_1^\star,
\end{equation}
by the positivity of $G_{rm}$. Since $(1-z^2)C(z) = 0$ at $z=-1$ we
deduce $(1 - z^2)C(z)>0$ for $-1<z<1$ and therefore $C(z)\geq 0$ for
$-1\leq z \leq 1$.

\section{Green's Functions and Related Quantities}\label{app:green}

In this section we collect several results regarding the following
three Green's functions for the unit ball in $\mathbb{R}^3$. First we
have \textit{membrane Green's function} $G_m(x,x_0)$ satisfying
\begin{equation}\label{eq:G_m_pde}
\Delta_{\partial\Omega} G_m = \frac{1}{|\partial\Omega|} -\delta_{\partial\Omega}(x-x_0),\quad x,x_0\in\partial\Omega;\qquad \int_{\partial\Omega} G_m dA = 0.
\end{equation}
Second, we have the \textit{Robin bulk Green's function} $G_{rb}(x,x_0)$ which solves
\begin{equation}\label{eq:G_rb_pde}
\Delta G_{rb} - \mu^2 G_{rb} = -\delta(x-x_0),\quad x,x_0\in \Omega;\qquad \partial_n G_{rb} + \kappa G_{rb} = 0,\quad x\in\partial\Omega,
\end{equation}
which is closely related to the \textit{Robin membrane Green's
  function} $G_{rm}(x,x_0)$ satisfying
\begin{equation}\label{eq:G_rm_greens_pde}
\Delta G_{rm} - \mu^2 G_{rm} = 0,\quad x\in\Omega;\qquad \partial_n G_{rm} + \kappa G_{rm} = \delta_{\partial\Omega}(x-x_0),\quad x,x_0\in\partial\Omega.
\end{equation}

\subsection{Series Expansions of Green's Functions}

The membrane Green's function is explicitly given by
\begin{subequations}
\begin{equation}\label{eq:green-membrane-exact}
G_m(x,x_0) = -\frac{1}{2\pi}\log|x-x_0| + R,\qquad R:=\frac{1}{4\pi}(\log 4 - 1),
\end{equation}
or in terms of a series expansion
\begin{equation}\label{eq:G_m_series}
G_m(x,x_0) = \frac{1}{4\pi} \sum_{l=1}^\infty \frac{2l+1}{l(l+1)}P_l(x\cdot x_0).
\end{equation}
\end{subequations}
The  free-space Green's function $G_f(x,x_0)$ satisfying
$$
\Delta G_f - \mu^2 G_f = -\delta(x-x_0),\quad x,x_0\in\mathbb{R}^3,\qquad G_f \rightarrow 0,\quad \text{as }|x-x_0|\rightarrow\infty,
$$
is explicitly given by
\begin{subequations}
\begin{equation}\label{eq:G_f_exact}
G_f(x,x_0) = \frac{e^{-\mu|x-x_0|}}{4\pi |x- x_0|}.
\end{equation}
Equivalently, it may be given in terms of an eigenfunction expansion as 
\begin{equation}\label{eq:G_f_series}
G_f(x,x_0) = \frac{1}{4\pi\sqrt{|x||x_0|}}\sum_{l=0}^\infty (2l+1)P_l\biggl(\frac{x\cdot x_0}{|x||x_0|}\biggr)\begin{cases} I_{l+1/2}(\mu |x|)K_{l+1/2}(\mu |x_0|),& |x|\leq |x_0|,\\ I_{l+1/2}(\mu |x_0|) K_{l+1/2}(\mu |x|),& |x| > |x_0|.\end{cases}
\end{equation}
\end{subequations}
Since $G_{rb}(x,x_0)$ has the same singularity as $G_f(x,x_0)$ we
decompose it as $G_{rb}(x,x_0) = G_f(x,x_0) + H_{rb}(x,x_0)$ and
calculate $H_{rb}(x,x_0)$ using an eigenfunction expansions. The
resulting series expansion for $G_{rb}(x,x_0)$ is
\begin{equation}\label{eq:G_rb_series}
G_{rb}(x,x_0) = \frac{1}{4\pi}\biggl\{\frac{e^{-\mu|x-x_0|}}{|x-x_0|} + \sum_{l=0}^{\infty}(2l+1)\tfrac{\mu K_{l+3/2}(\mu) - (l+\kappa) K_{l+1/2}(\mu)}{\mu I_{l+3/2}(\mu) + (l+\kappa) I_{l+1/2}(\mu)}\tfrac{I_{l+1/2}(\mu |x|)I_{l+1/2}(\mu |x_0|)}{\sqrt{|x||x_0|}}P_l\biggl(\frac{x\cdot x_0}{|x||x_0|}\biggr)\biggr\},
\end{equation}
where $|x_0|< 1$. Note that if $x_0 = 0$ then only the $l=0$ term
remains. The series expansion for the Robin membrane Green's function
is found to be
\begin{equation}\label{eq:G_rm_series}
G_{rm}(x,x_0) = \frac{1}{4\pi}\sum_{l=0}^{\infty}g_l(|x|) P_l\biggl(\frac{x\cdot x_0}{|x|}\biggr),
\end{equation}
where
\begin{equation}\label{eq:series-gl}
g_l(z) := \frac{2l+1}{\mu I_{l+3/2}(\mu) + (\kappa+l)I_{l+1/2}(\mu)}\frac{I_{l+1/2}(\mu z)}{\sqrt{z}},\qquad z\geq 0,
\end{equation}
and we make note of the special case
$$
g_l(0) = \begin{cases} \sqrt{\frac{2\mu}{\pi}}\frac{1}{\mu I_{3/2}(\mu) + \kappa I_{1/2}(\mu)} & l = 0, \\ 0 & l\geq 1,\end{cases}
$$
obtained by using the well known asymptotics
$I_\nu(z) \sim \tfrac{1}{2^\nu \Gamma(\nu+1)}z^\nu $ as
$z\rightarrow 0^+$. Note that by using the series for $G_{rb}$ and
that for $G_f$ as well as the differentiation and Wronskian identities
$$
I_{\nu}'(\mu) = \frac{\nu}{\mu}I_{\nu}(\mu) + I_{\nu+1}(\mu) ,\quad K_{\nu}'(\mu) = \frac{\nu}{\mu}K_{\nu}(\mu) - K_{\nu+1}(\mu),\quad K_{\nu}(\mu) I_{\nu}'(\mu) - K_{\nu}'(\mu) I_{\nu}(\mu) = \frac{1}{\mu},
$$
we calculate that for any $x\in\partial\Omega$ and $y\in\Omega$ the
following reciprocity formula holds
\begin{equation}\label{eq:G_rb_on_membrane}
G_{rb}(x,y) = G_{rm}(y,x) = \frac{1}{4\pi}\sum_{l=0}^{\infty}g_l(|y|) P_l\biggl(\frac{x\cdot y}{|y|}\biggr).
\end{equation}

Finally we make note of the following two formulas
\begin{equation}\label{eq:G_rm-and-G_rb-integrals}
\int_{\partial\Omega} G_{rm}(x,y) dA_x = g_0(1),\quad\text{for any } y\in\partial\Omega,\quad\text{and}\quad \int_{\partial\Omega} G_{rb}(x,y) dA_x = g_0(y)\quad\text{for any } y\in\Omega.
\end{equation}

\subsection{Surface Integrals of Product with $G_m$}

The next two identities will be useful in this and the following sections. Let 
$$
x_i = (\sin\theta_i\cos\varphi_i, \sin\theta_i\sin\varphi_i,\cos\theta_i)^T
$$
and recall the summation formula for Legendre polynomials
\begin{subequations}
\begin{equation}\label{eq:legendre_summation}
P_l(x_i^T x_j) = \sum_{m=-l}^l \frac{(l-m)!}{(l+m)!} P_l^m(\cos\theta_i) P_l^m(\cos\theta_j)e^{im(\varphi_i - \varphi_j)},
\end{equation}
which we use to calculate
\begin{equation}\label{eq:legendre_product}
\int_{\partial\Omega} P_l(x_i^T x)P_k(x^T x_j) dA_x = \delta_{kl}\frac{4\pi}{2l+1}P_l(x_i^T x_j).
\end{equation}
\end{subequations}
Next we define
\begin{equation}\label{eq:definition-I_Gm}
I_{G_m}(r,z) = \frac{1}{4\pi}\sum_{l=1}^\infty \frac{g_l(r)}{l(l+1)} P_l(z),\qquad\text{for any }0\leq r\leq 1,\quad -1\leq z\leq 1,
\end{equation}
so that by using the series expansions \eqref{eq:G_m_series},
\eqref{eq:G_rb_series}, \eqref{eq:G_rm_series}, and the product
formula \eqref{eq:legendre_product} we calculate
\begin{subequations}
\begin{align}
& \int_{\partial\Omega} G_m(x_i,x)G_{rb}(x,x_0)dA_x = I_{G_m}(|x_0|,x_i^Tx_0/|x_0|), & \text{for any }x_i\in\partial\Omega\text{ and } x_0\in\Omega, \\
& \int_{\partial\Omega} G_m(x_i,x) G_{rm}(x,x_j) dA_x = I_{G_m}(1,x_i^Tx_j), & \text{for any }x_i,x_j\in\partial\Omega
\end{align}
\end{subequations}

\subsection{Surface Integrals of Products with $\nabla_{\mathbb{R}^3} G_m$}

In this section we will derive useful computational formulae for the evaluation of
$$
I_{\nabla G_m}(x_0,x_1) = \int_{\partial\Omega} \frac{x_0 - x}{|x_0 - x|^2} f(x,x_1) dA_x,
$$
for any $x_0\in\partial\Omega$ and $x_1\in\Omega\cup\partial\Omega$ where we assume
$$
f(x,x_1) = \frac{1}{4\pi}\sum_{l=0}^\infty f_l P_l\biggl(\frac{x\cdot x_1}{|x_1|}\biggr),
$$
and allow $f_l = f_l(|x|,|x_1|)$. Let $\mathcal{R}_0$ be a rotation matrix such that
$\mathcal{R}_0 x_0 = e_z = (0,0,1)^T$. Then
$$
I_{\nabla G_m}(x_0,x_1) = \mathcal{R}_0^T \int_{\partial\Omega}\frac{\mathcal{R}_0 x_0 - \mathcal{R}_0 x}{|\mathcal{R}_0 x_0 - \mathcal{R}_0 x|^2} f(\mathcal{R}_0^T \mathcal{R}_0 x, x_1) dA_x = \mathcal{R}_0^T \int_0^\pi \int_0^{2\pi} {\footnotesize \begin{pmatrix} -\sin\theta\cos\varphi \\ -\sin\theta\sin\varphi \\ 1 - \cos\theta \end{pmatrix}}\frac{f(x,\tilde{x}_1)}{2-2\cos\theta}\sin\theta d\theta d\varphi,
$$
where $\tilde{x}_1 = \mathcal{R}_0 x_1$. The $z$ component of the integral is easily calculated to be
\begin{equation*}
\frac{1}{2}\int_{\partial\Omega} f(x,\tilde{x}_1) dA_x = \frac{1}{2} f_0.
\end{equation*}
Note that if $x_1$ is collinear with $x_0$ then in the integral $f(x,\tilde{x}_1)$ is a function only of $\theta$ and therefore the $x$ and $y$ components of the integral vanish, leaving only the $z$ component calculated above. For the remainder of the calculation we therefore assume that $x_1$ and $x_0$ are not collinear. In this case the $x$ and $y$ components can be obtained as the real and \textit{negative}
imaginary parts of
$$
J = -\frac{1}{2}\int_0^\pi \int_0^{2\pi}\frac{e^{-i\varphi} \sin\theta}{1-\cos\theta}f(x,\tilde{x}_1)\sin\theta d\theta d\varphi,
$$
which we calculate to be 
\begin{align*}
J & = -\frac{1}{8\pi}\sum_{l=0}^\infty f_l \sum_{m=-l}^l \tfrac{(l-m)!}{(l+m)!} P_l^m(\cos\tilde{\theta}_1)e^{-im\tilde{\varphi}_1}\int_0^{2\pi}e^{i(m-1)\varphi}d\varphi\int_{-1}^1 \sqrt{\frac{1+x}{1-x}}P_l^m(x)dx \\
& = -\frac{1}{4} \sum_{l=1}^\infty \frac{f_l}{l(l+1)} P_l^1(\cos\tilde{\theta}_1)e^{-i\tilde{\varphi}_1} \int_{-1}^1 \sqrt{\frac{1+x}{1-x}}P_l^1(x)dx  = \frac{1}{2} \sum_{l=1}^\infty \frac{f_l}{l(l+1)} P_l^1(\cos\tilde{\theta}_1)e^{-i\tilde{\varphi}_1},
\end{align*}
where the last equality is obtained by using $P_l^1(x) = -\sqrt{1-x^2}P_l'(x)$ and the normalization $P_l(1)=1$ to calculate for $l\geq 1$
$$
\int_{-1}^{1} \sqrt{\frac{1+x}{1-x}}P_l^1(x)dx = -\int_{-1}^1 (1+x) P_l'(x) dx = -(1+x)P_l(x)\bigr|_{-1}^1 + \int_{-1}^1 P_l(x)dx = -2P_l(1) = -2.
$$
Therefore we calculate
$$
I_{\nabla G_m}(x_0,x_1) = \frac{1}{2}\mathcal{R}_0^T\begin{pmatrix} \cos\tilde{\varphi}_1 \\ \sin\tilde{\varphi}_1 \\ 0 \end{pmatrix} \sum_{l=1}^\infty \frac{f_l}{l(l+1)}P_l^1(\cos\tilde{\theta}_1) + \frac{1}{2}\mathcal{R}_0^T e_z f_0.
$$
We make some further simplifications. First, notice that
$$
\cos\tilde{\theta_1} = \frac{e_z^T\tilde{x}_1}{|\tilde{x}_1|} = e_z^T\mathcal{R}_0 \frac{x_1}{|x_1|} = (\mathcal{R}_0^T e_z)^T \frac{x_1}{|x_1|} = \frac{x_0\cdot x_1}{|x_1|}.
$$
Second, we note that since we are assuming that $x_0$ and $x_1$ are
not colinear we must have $0<\tilde{\theta_1}<\pi$ so that
$\sin\tilde{\theta_1}\neq 0$ and the first vector in the expression
for $I_{\nabla G_m}(x_0,x_1)$ therefore simplifies to
$$
\mathcal{R}_0^T\begin{pmatrix} \cos\tilde{\varphi}_1 \\ \sin\tilde{\varphi}_1 \\ 0 \end{pmatrix} = \frac{1}{|x_1|}\biggl(\frac{1}{\sin\tilde{\theta}_1}\mathcal{R}_0^T \tilde{x}_1 - \cot\tilde{\theta_1}\mathcal{R}_0^T e_z\biggr) = \frac{\mathbb{I} - x_0x_0^T}{\sqrt{|x_1|^2-(x_0\cdot x_1)^2}} x_1,
$$
which is the normalized projection of $x_1$ to the plane
orthogonal to $\partial\Omega$ at $x_0$. Summarizing we obtain the formula
\begin{equation*}
\int_{\partial\Omega} \frac{x_0 - x}{|x_0 - x|^2} f(x,x_1) dA_x = \frac{1}{2} f_0 x_0 + \frac{1}{2}\sum_{l=1}^\infty \frac{f_l}{l(l+1)}P_l^1\biggl(\frac{x_0\cdot x_1}{|x_1|}\biggr)\frac{\mathbb{I} - x_0x_0^T}{\sqrt{|x_1|^2-(x_0\cdot x_1)^2}}x_1,
\end{equation*}
for any $x_0\in\partial\Omega$ and $x_1\in\Omega\cup\partial\Omega$.

By the preceding computations, if we define
\begin{subequations}
\begin{align}
I_{\nabla G_m}^\parallel(r) & = \frac{1}{2}g_0(r),\label{eq:I_grad_parallel}\\
I^{\perp}_{\nabla G_m}(r,z) & = \frac{1}{2}\sum_{l=1}^\infty \frac{g_l(r)}{l(l+1)}P_l^1(z),\label{eq:I_grad_perp}
\end{align}
\end{subequations}
then for any $x_i,x_j\in\partial\Omega$ and $x_0\in\Omega$ we have
\begin{subequations}
\begin{align}
& \int_{\partial\Omega}\frac{x_i-x}{|x_i-x|^2} G_{rm}(x,x_j) dA_x = I_{\nabla G_m}^\parallel(1)x_i + I^{\perp}_{\nabla G_m}(1,x_i^Tx_j)\frac{\mathbb{I} - x_ix_i^T}{\sqrt{1-(x_i^T x_j)^2}}x_j, \label{eq:integral_grad_Gm_Grm}\\
& \int_{\partial\Omega}\frac{x_i-x}{|x_i-x|^2} G_{rb}(x,x_0) dA_x = I_{\nabla G_m}^\parallel(|x_0|)x_i + I^{\perp}_{\nabla G_m}(1,x_i^T\hat{x}_0)\frac{\mathbb{I} - x_ix_i^T}{\sqrt{1-(x_i^T \hat{x}_0)^2}}\hat{x}_0, \label{eq:integral_grad_Gm_Grb}
\end{align}
\end{subequations}
where $\hat{x}_0 := x_0/|x_0|$.

We conclude by making note of the useful identity
\begin{equation}\label{eq:I_Gm_derivative}
\frac{\partial I_{G_m}(\eta_0,\cos\theta)}{\partial\theta} = \frac{1}{2\pi}I_{\nabla G_m}^\perp(\eta_0,\cos\theta).
\end{equation}

 \addcontentsline{toc}{section}{References}
 \bibliographystyle{abbrv}
 \bibliography{bruss_sphere_bibliography}

\end{document}